\documentclass[aps,prb,twocolumn,amsmath,amssymb,nofootinbib,superscriptaddress,floatfix,epsfig]{revtex4-1}

\usepackage{amsmath,cleveref,amssymb,amsfonts,graphicx,multirow,color,subfigure,slashbox,bm,textcomp,ctable}

\usepackage{srcltx,soul}
\usepackage[usenames,dvipsnames,svgnames,table]{xcolor}

\crefname{equation}{Eq.}{Eqs.}

\newcommand{\bra}[1]{\langle #1 |}
\newcommand{\ket}[1]{| #1 \rangle}
\newcommand{\braket}[2]{\left \langle #1 | #2 \right\rangle}

\newcommand{\be}{\begin{equation}}
\newcommand{\ee}{\end{equation}}
\newcommand{\bma}{\begin{pmatrix}}
\newcommand{\ema}{\end{pmatrix}}
\newcommand{\balig}{\begin{align}}
\newcommand{\ealig}{\end{align}}

\newcommand{\bZ}{\mathbb{Z}}
\newcommand{\bI}{\mathbb{I}}
\newcommand{\ba}{\begin{eqnarray}}
\newcommand{\ea}{\end{eqnarray}}

\newcommand{\ignore}[1]{}

\newcommand{\CTR}{[T,R]=0}
\newcommand{\CCR}{[C,R]=0}
\newcommand{\ATR}{\{T,R\}=0}
\newcommand{\ACR}{\{C,R\}=0}
\newcommand{\CSR}{[S,R]=0}
\newcommand{\ASR}{\{S,R\}=0}
\newcommand{\mH}{\mathcal{H}}
\newcommand{\mM}{\frak{M}}
\newcommand{\six}{\sigma_x}
\newcommand{\siy}{\sigma_y}
\newcommand{\siz}{\sigma_z}
\newcommand{\mN}{\frak{N}}

\begin{document}

\title{
    Classification of topological insulators and superconductors
    in the presence of reflection symmetry
}

\author{Ching-Kai Chiu}
\affiliation{Department of Physics, University of Illinois, Urbana, IL 61801, USA}

\author{Hong Yao}
\affiliation{
Institute for Advanced Study, Tsinghua University, Beijing, 100084, China
}

\author{Shinsei Ryu}
\affiliation{Department of Physics, University of Illinois, Urbana, IL 61801}

\date{\today}

\begin{abstract}
We discuss a topological classification of
insulators and superconductors
  in the presence of both (non-spatial) discrete symmetries
in the Altland-Zirnbauer classification and
spatial reflection symmetry in any spatial dimensions.
By using the structure of bulk Dirac Hamiltonians
of minimal matrix dimensions and explicit constructions of
topological invariants,
we provide the complete classification, which still has the same dimensional periodicities with the original Altland-Zirnbauer classification.
The classification of reflection-symmetry-protected topological insulators and superconductors
depends crucially on the way reflection symmetry operation is realized.
When a boundary is introduced, which is reflected into itself,
these non-trivial topological insulators and superconductors support gapless modes localized
at the boundary.
\end{abstract}


\maketitle



\section{Introduction}

Topological insulators and superconductors are
symmetry preserving fermionic systems with a bulk energy gap\cite{hasan:rmp,qi:rmp}.
Relevant symmetry conditions that are necessary to define these symmetry protected topological states can be
divided into two categories: \emph{non-spatial} symmetries and \emph{spatial} symmetries.
The Hamiltonians of \emph{non-spatial} symmetric systems
may possess time reversal symmetry (TRS), particle-hole symmetry (PHS),
or chiral symmetry. They have gapless boundary states that are topologically protected
and are related to physical quantities, such as the Hall conductivity.
The subject started with the recognition by Kane and Mele
\cite{Kane:2005kx, Kane:2005vn}
that by incorporating a spin-orbit coupling in the tight-binding model for graphene, the system 
will become what is now known as a 2D $\mathbb{Z}_2$ topological insulator with time reversal symmetry (TRS), 
as well as the theoretical prediction\cite{Bernevig:2006kx} and experimental observation\cite{MarkusKonig11022007} of such $\mathbb{Z}_2$ topological states in the HgTe/CdTe quantum well.
After that,
three-dimensional (3D)
$\mathbb{Z}_2$
topological insulators were predicted
\cite{Fu:2007fk,Moore:2007uq,Roy:2009kx}
and observed;\cite{Hsieh:2008fk,Chen:2009vn,D.Hsieh02132009,Xia:2009uq}
the identification of $^{3}$He-B as a topological superconductor
was realized.\cite{Volovik:book,Wada:2008ly,Murakawa:2009ve,JPSJ.80.013602}
It turns out that those topological insulators and superconductors
are just a part of a larger scheme ---
a complete classification of topological insulators and superconductors
has been developed 
with a unified periodic table \cite{Schnyder:2008gf,Kitaev,qi:2008sf}. 


Topological phases protected by these {\it non-spatial} discrete symmetries
are stable against spatially homogeneous as well as inhomogeneous deformations.
In addition, the protected boundary modes (edge, surface, etc)
that appear
at the boundary of topological phases in the periodic table
are completely immune to disorder;
for arbitrary strong disorder, as far as the bulk topological character is
not altered in the bulk through a phase transition,
the boundary can never be Anderson-localized.\cite{Schnyder:2008gf}


With a set of discrete {\it spatial} symmetries, a topological distinction among gapped phases
(\emph{i.e.,}
``symmetry protected topological phase'')
can arise as well. 
One example is 
inversion symmetry protected topological 
insulators\cite{Turner:2010qf,Turner:2012bh, Hughes:2011uq} 
where inversion symmetry is defined as the invariance of the system
under the sign flip of the spatial coordinate
$r\to -r$,
where $r=(r_1,r_2,\ldots, r_d)$ is the spatial coordinates
in $d$ spatial dimensions, $r \in \mathbb{R}^d$
(for lattice systems, $r$ labels a site
on a $d$-dimensional lattice).
Unlike the case of non-spatial discrete symmetries,
for
topological phases protected by a set of spatial symmetries,
non-trivial bulk topology
is not necessarily accompanied by a gapless boundary mode,
as 
the boundary 
might break the spatial symmetries in question. The implications of certain specific point group symmetries
on the topological distinction of ground states
has been also discussed.\cite{Fu:2011cr,Slager:2013fk,Fang:2012kx,Fang:2012uq} More importantly, Hsieh {\it et al.\/}\cite{Hsieh:2012fk} predicted that one of such \emph{spatial} symmetric insulators can be realized in the Pb$_{1-x}$Sn$_x$Te material class, and this prediction was experimentally verified by ARPES.\cite{Xu:2012kx}

\begin{figure}
\includegraphics[width=50mm]{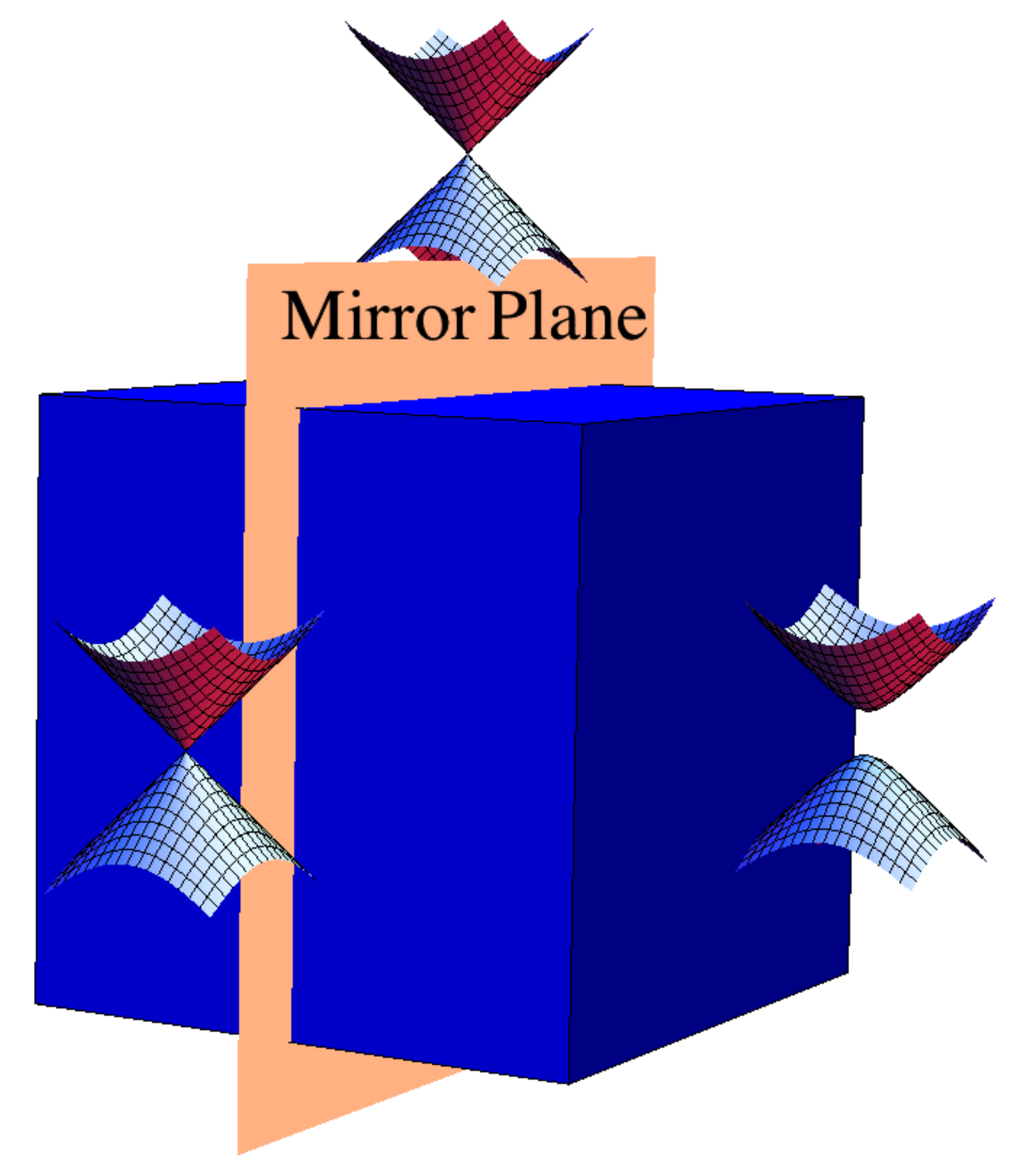}
\caption{
 For topological insulators and superconductors
 that are protected by reflection symmetry,
 the correspondence between gapless surface states and bulk topology
 holds only when the surface that reflects to itself is chosen.
 The figures shows that nontrivial bulk topology guarantees gapless states
 in the \emph{self-reflected} surfaces.
 Furthermore, gapped states in the \emph{non-self-reflected} surfaces does not imply trivial bulk topology.  }
\label{reflectioncube}
\end{figure}

In this paper, we discuss the implication of a reflection (or mirror) symmetry in one spatial direction:
it is an invariance of the system under the sign flip of, say, the first component of Cartesian coordinates,
$r\to \tilde{r}\equiv (-r_1,r_2,\ldots, r_d)$. (For an earlier study of this subject, see Ref.\ \onlinecite{Teo:2008fk}.) While an inversion symmetry singles out a special point,
reflection symmetry singles out a special $(d-1)$-dimensional plane
($r_1=0$ in the this case).
As a consequence,
when we terminate the system with a
$(d-1)$-dimensional boundary (plane) which is orthogonal to
the reflection plane ($r_1=0$),
the boundary with constant $r_i$ ($i\neq 1$) is reflection symmetric
under $(r_1,r_2,\ldots, r_d) \to (-r_1,r_2,\ldots,r_d)
$. Reflection symmetry is arguably the simplest spatial symmetry of a system for which certain boundaries can respect the spatial symmetry in question.
This boundary property is an important distinction
from the inversion symmetric topological
phases, for which a plane boundary to the system alone does not inherit
the spatial symmetry (inversion symmetry) in the bulk.
(A pair of boundaries can be inversion symmetric to each other, though).
With the special choice of the boundary above,
we will argue that for topologically non-trivial phases
protected by reflection
there is a stable boundary mode,
in the manner similar to
topologically phases protected by non-spatial discrete symmetries. The correspondence still holds between the non-trivial bulk topology and the gapless boundary modes when the boundary that reflects to itself is chosen as shown in \Cref{reflectioncube}.

Although topological insulators and superconductors
protected by \emph{non-spatial} or \emph{spatial} symmetries have been studied separately,
their recognition does not directly provide a complete classification of topological systems
in the presence of both non-spatial and spatial discrete symmetries. 
Therefore, we will consider the topological classification of reflection symmetric systems
with a subset of the three non-spatial symmetries: TRS, PHS, and chiral symmetry.
We found the topological classification depends not only on
the set of symmetries which are respected 
but also on the way how reflection symmetry is realized,
i.e., algebraic relations satisfied
among reflection and other non-spatial discrete symmetries
when they exist.
Our result of the classification of reflection symmetric systems
is summarized in \Cref{table:mirror class}.
The algebraic relations among reflection and
non-spatial symmetry operations are denoted by
$R_{\pm}$ and $R_{\pm,\pm}$ in \Cref{table:mirror class}.

Non-trivial topological states displayed
in \Cref{table:mirror class} are characterized by
a topological invariant of integer ($\mathbb{Z}_2$) or $\mathbb{Z}_2$ type.
For example,
the entries in \Cref{table:mirror class}
marked by $M\mathbb{Z}$
indicate the presence of topologically protected states by
a topological invariant
defined on mirror invariant planes in the Brillouin zone
(``mirror topological invariant'').
These topological states
include
3D topological insulators
protected by the ``mirror Chern number''
discussed in Ref.\ \onlinecite{Teo:2008fk},
and 2D topological superconductors with TRS and reflection symmetry
(class DIII $+R$) discussed in Ref.\ \onlinecite{Yao:2012},
which are characterized by reflection winding numbers in the
1D mirror lines.
We will generalize those mirror numbers to any spatial dimensions
and relevant symmetry classes.
Furthermore,
we will show that some systems are protected by the $\bZ$ topological number and the mirror Chern number simultaneously;
the larger one of these two numbers
gives
a new integral topological invariant (denoted by $\bZ^1$
in \Cref{table:mirror class}).
In other cases, the topolgoical insulators and superconductors in the original periodic table
turn out to be invariant under reflection.
If this is the case, the same topological invariant also characterizes the
non-trivial topology of reflection-symmetric topological insulators and superconductors.
These cases are indicated by ``0'', $\mathbb{Z}_2$, and $\mathbb{Z}$
in \Cref{table:mirror class}.
In short, we claim that the reflection-symmetric topological states are characterized by one
of the topological invariants,
`$0$', $\bZ_2$, $\bZ$, $M\mathbb{Z}$, and $\mathbb{Z}^1$.

In this manuscript,
we use ``the minimal Dirac Hamiltonian method''
to characterize the AZ symmetry classes with reflection symmetry.
Without reflection symmetry the topological classification (\Cref{originaltable})
of the AZ symmetry classes can be studied by
Anderson localization,\cite{Schnyder:2008gf}
K-theory,\cite{Kitaev}
and minimal Dirac Hamiltonians.\cite{Stone:2011qo,unpublish3rd}
The minimal Dirac Hamiltonian method provides a direct way to produce the original classification (\Cref{originaltable}).
In this method,
we first write down
a bulk Dirac Hamiltonian preserving system's symmetries in the minimal matrix dimension.
The topological class of the system is determined by the existence of a symmetry preserving
extra mass term (SPEMT), which keeps the system in the same topological phase during the continuous deformation.
If this term exists in the minimal Dirac Hamiltonian,
this phase is characterized by `0' topological invariant.
If not, we consider a bigger system including two minimal Dirac Hamiltonians.
The presence of a SPEMT in this system of the two copies implies $\bZ_2$ topological invariant character.
Otherwise, the absence of a SPEMT implies $\bZ$ character.
When classifying reflection symmetry topological insulators and superconductors,
we study the existence of a SPEMT in Dirac Hamiltonians to determine topological characters.
 Complementary to
the minimal Dirac Hamiltonian method,
we also look for topological invariants ($0,\ \bZ_2,\ \bZ,\ M\bZ$ and $\bZ^1$)
in the presence of reflection symmetry to determine bulk topology.
The classification of bulk topology in terms of topological invariants
is fully consistent
with the minimal Dirac Hamiltonian method.



Topological insulators might have topological invariants of strong and weak indices.\cite{Fu:2007uq} In the original classification table\cite{Kitaev,Schnyder:2008gf} for a $d$-dimensional system, the strong index is the topological invariant in $d$ dimensions and the weak indices are captured by the strong indices in the dimensions less than $d$. However, the complications arise when the weak indices are considered in the classification of reflection symmetric systems. The reason is that the weak indices may not be described by the strong indices in the dimensions less than $d$ in the reflection classification table. Moreover, the weak indices might depend on spatial directions. That is, in different directions the weak indices are different because reflection symmetry operation only flips one direction. In this manuscript, we shall focus on only the strong indices of the classification.

To name a few physically interesting
topological systems
in \Cref{table:mirror class},
in symmetry class AII in three dimensions,
with reflection symmetry specified by $R_{--}$,
there are topological insulators protected by
the $M\bZ$ invariant.
These are nothing but the topological insulator
that was proposed by Hsieh \emph{et al.}\cite{Hsieh:2012fk}
and observed by Xu \emph{et al.}\cite{Xu:2012kx} in the Pb$_{1-x}$Sn$_x$Te material class.
It is the first experimental realization of crystalline topological insulators.
This particular reflection symmetric topological insulator
continues to be topologically non-trivial even in the absence of
TRS,
as indicated by
``$M\mathbb{Z}$''
in symmetry class A in three dimensions
in \Cref{table:mirror class}.
\Cref{table:mirror class} also includes
topological superconductors protected by reflection symmetry,
such as with TRS and reflection symmetry
2D topological superconductors
(class DIII $+R_{--}$),\cite{Yao:2012} which are classified by an integral-valued topological invariant.
For symmetry class D in two dimensions, which hosts T-breaking
topological superconductors in the absence of reflection symmetry,
there is a reflection symmetric topological superconductor
characterized by a $\mathbb{Z}_2$
topological invariant.
Other examples will also be discussed in Sec.\ \ref{examples}.

The manuscript is organized as follows:
in Sec.\ \ref{reflectionband},
we provide the background knowledge of reflection symmetry in band theory,
and in particular, describe how we distinguish different realizations
of reflection symmetry operation in the presence of
other non-spatial discrete symmetries.
In Sec.\ \ref{DiracHam},
we review the connection between the minimal bulk Dirac Hamiltonians
and the topology possessing `$0$', $\bZ_2$ and $\bZ$ topological invariants
in the Altland-Zirnbauer(AZ) symmetry classes
without reflection symmetry.
Moreover, by considering reflection-symmetric Dirac Hamiltonians we show the correspondence between gapless boundary states and bulk topology.
In Sec.\ \ref{shifttable}, we consider the two kinds of the classifications:
the reflection symmetry operator commutes with all of the non-spatial discrete symmetries and anticommutes with TRS and PHS operators.
In Sec.\ \ref{restclassification}, we classify the remaining of the AZ symmetry classes possessing TRS and PHS under the condition that one of these symmetry operators commutes with the reflection symmetry operator and the other anticommutes with the reflection symmetry operator.
In Sec.\ \ref{examples}, the concrete examples for
topological insulators/superconductors
protected by reflection symmetry are provided.

\begin{table*}[t]
\begin{center}
\begin{tabular}{ |c | c  c  c || c | c||c|c|c|c|c|c|c|c|}\hline
  ~~ AZ Class~~  &   ~~T~~  & ~~C~~ & ~~S~~  &  ~~$R$ operator~~   & ~~MSC~~ & ~$d=1$~ & ~$d=2$~ & ~$d=3$~ & ~$d=4$~ & ~$d=5$~ & ~$d=6$~ & ~$d=7$~ & ~$d=8$~ \\
\hline
\hline
\multirow{2}{*}{AIII} &   \multirow{2}{*}{0} & \multirow{2}{*}{0} &\multirow{2}{*}{1} &  $R_+$   &  AIII$^2$ & 0 & $M\bZ$ & 0 & $M\bZ$ & 0 & $M\bZ$ & 0 & $M\bZ$  \\
  \cline{5-14}
   & &  &  &  $R_-$ & A & $\bZ^1$ & 0 & $\bZ^1$ & 0 & $\bZ^1$ & 0 & $\bZ^1$ & 0 \\
\hline
A &   0 & 0 & 0  						&   $R$  & A$^2$ & $M\bZ$ & 0 & $M\bZ$ & 0 & $M\bZ$ & 0 & $M\bZ$ & 0 \\
\hline	
\hline		
   \multirow{2}{*}{AI} &   \multirow{2}{*}{+} & \multirow{2}{*}{0}  & \multirow{2}{*}{0}  					& $R_{+}$\footnotemark[3]   & AI$^2$ & $M\bZ$ & 0 & 0 & 0 & $2M\bZ$ & 0 & $\bZ_2$ & $\bZ_2$   \\
      \cline{5-14}
   & &  &  &  $R_{-}$ &    A  & 0 & 0 & $2M\bZ$ & 0 & 0 & $\mathbb{Z}_2$ & $M\bZ$ & 0 \\
\hline
   \multirow{4}{*}{BDI} &    \multirow{4}{*}{+}  & \multirow{4}{*}{+} & \multirow{4}{*}{1}   		&  $R_{++}$\footnotemark[3]     & BDI$^2$ & $\mathbb{Z}_2$ & $M\bZ$ & 0 & 0 & 0 & $2M\bZ$ & 0 & $\bZ_2$   \\
   \cline{5-14}
   & &  &  &  $R_{--}$  & AIII & 0 & 0 & 0 & $2M\bZ$ & 0 & 0  & $\bZ_2$ & $M\bZ$ \\
      \cline{5-14}
   & &  &  &  $R_{+-}$  & AI  &   $2\bZ^1$ & $0$ & 0 & 0 & $\bZ^1$ & 0 & $\bZ_2$ & $\bZ_2$ \\
      \cline{5-14}
   & &  &  &  $R_{-+}$   & D &  $2\bZ$  & $0$ & $2M\bZ$ & 0 & $2\bZ$ & $0$ & $2M\bZ$ & 0   \\
\hline
  \multirow{2}{*}{D} &   \multirow{2}{*}{0} & \multirow{2}{*}{+} & \multirow{2}{*}{0}   	 						& $R_{+}$\footnotemark[3]  & D$^2$ & $\mathbb{Z}_2$ & $\mathbb{Z}_2$ & $M\bZ$ & 0 & 0 & 0 & $2M\bZ$ & 0 \\
  \cline{5-14}
   & &  &  &  $R_{-}$\footnotemark[4]  & A & $M\bZ$ & 0 & 0 & 0 & $2M\bZ$ & 0  & 0 & $\bZ_2$ \\
\hline
  \multirow{4}{*}{DIII} &    \multirow{4}{*}{-} & \multirow{4}{*}{+} & \multirow{4}{*}{1}  					
  & $R_{++}$  &  DIII$^2$  & 0 & $\mathbb{Z}_2$ & $\mathbb{Z}_2$ & $M\bZ$ & 0 & 0 & 0 & $2M\bZ$ \\
\cline{5-14}
   & &  &  &  $R_{--}$\footnotemark[4] &   AIII & $\mathbb{Z}_2$ & $M\bZ$ & 0 & 0 & 0 & $2M\bZ$ & 0 & 0  \\
   \cline{5-14}
   & &  &  &  $R_{+-}$ &  AII & $2M\bZ$ & 0 & $2\bZ$ & $0$ & $2M\bZ$ & 0  & 0 & $2\bZ$   \\
   \cline{5-14}
   & &  &  &  $R_{-+}$ &    D &  $\bZ_2$ & $\bZ_2$ & $\bZ^1$ & 0 & 0 & 0 & $2\bZ^1$ & 0 \\
\hline
  \multirow{2}{*}{AII} &   \multirow{2}{*}{-} &  \multirow{2}{*}{0}  	 & \multirow{2}{*}{0}					& $R_{+}$ &  AII$^2$ & $2M\bZ$ & 0 & $\mathbb{Z}_2$ & $\mathbb{Z}_2$ & $M\bZ$ & 0 & 0 & 0 \\
      \cline{5-14}
   & &  &  &  $R_{-}$\footnotemark[4] &    A & 0 & $\mathbb{Z}_2$ & $M\bZ$ & 0 & 0 & 0 & $2M\bZ$ & 0 \\
\hline
  \multirow{4}{*}{CII} &   \multirow{4}{*}{-} & \multirow{4}{*}{-} & \multirow{4}{*}{1} 	  		&  $R_{++}$ &   CII$^2$ & 0 & $2M\bZ$ & 0 & $\mathbb{Z}_2$ & $\mathbb{Z}_2$ & $M\bZ$ & 0 & 0 \\
  \cline{5-14}
   & &  &  &  $R_{--}$ &  AIII & 0 & 0 & $\mathbb{Z}_2$ & $M\bZ$ & 0 & 0   &0 & $2M\bZ$ \\
   \cline{5-14}
   & &  &  &  $R_{+-}$ & AII  &  $2\bZ^1$ & 0 & $\bZ_2$ & $\bZ_2$ & $\bZ^1$ & 0 & 0   & 0 \\
   \cline{5-14}
   & &  &  &  $R_{-+}$ &    C  &  $2\bZ$  & $0$ & $2M\bZ$ & 0 & $2\bZ$ & $0$ & $2M\bZ$ & 0   \\
\hline
  \multirow{2}{*}{C} &  \multirow{2}{*}{0}  & \multirow{2}{*}{-} & \multirow{2}{*}{0} 	 						&  $R_{+}$\footnotemark[5]  &  C$^2$& 0 & 0 & $2M\bZ$ & 0 & $\mathbb{Z}_2$ & $\mathbb{Z}_2$ & $M\bZ$ & 0     \\
    \cline{5-14}
   & &  &  &  $R_{-}$ &    A & $2M\bZ$ & 0 & 0 & $\mathbb{Z}_2$ & $M\bZ$ & 0 & 0 & 0   \\
\colrule
  \multirow{4}{*}{CI} &   \multirow{4}{*}{+}  & \multirow{4}{*}{-} & \multirow{4}{*}{1}  					&   $R_{++}$\footnotemark[6] &  CI$^2$ & 0 & 0 & 0 & $2M\bZ$ & 0 & $\mathbb{Z}_2$ & $\bZ_2$& $M\bZ$  \\
  \cline{5-14}
   & &  &  &  $R_{--}$ &  AIII & 0 & $2M\bZ$ & 0 & 0 & $\mathbb{Z}_2$ & $M\bZ$ & 0 &0   \\
   \cline{5-14}
   & &  &  &  $R_{+-}$ &    AI & $2M\bZ$ & 0 & $2\bZ$ & $0$ & $2M\bZ$ & 0  & 0 & $2\bZ$  \\
   \cline{5-14}
   & &  &  &  $R_{-+}$ &    C &  0&  $0$ & $2\bZ^1$ & 0 & $\bZ_2$ & $\bZ_2$ & $\bZ^1$ & 0\\
\hline
\end{tabular}
\footnotetext[3]{Spinless systems} \footnotetext[4]{Spin-$1/2$ systems}
\footnotetext[5]{Spin-$1/2$, $C^2=1$, $SU(2)$ symmetry for the spin. }
\footnotetext[6]{Spin-$1/2$, $C^2=1$ and
 $T^2=-1$
, $SU(2)$ symmetry for the spin.}
\caption{The complete classification table of reflection-symmetric topological insulators and superconductors: For class AIII, $R_{\pm}$ indicates that the reflection symmetry operator ($R$) commutes/anticommutes with $S$. For four real symmetry classes ($R_{\pm\pm},\ R_{\pm\mp}$) that have TRS and PHS, the first sign $\pm$ of $R$ indicates that $R$ commutes/anticommutes with $T$ and the second sign $\pm$ indicates that $R$ commutes/anticommutes with $C$. For the four other real symmetry classes ($R_{\pm}$) that preserve only one non-spatial symmetry, the sign $\pm$ indicates that $R$ commutes/anticommutes with system's non-spatial symmetry operator. The Hamiltonian in the mirror symmetry plane can be block-diagonalized to two blocks in the eigenspace $R=\pm 1$. The superscript of $2$ in the mirror symmetry classes (MSC) (See Appendix \ref{mirror class}) indicates that these two blocks of $R=\pm 1$ are independent. }
\label{table:mirror class}
\end{center}
\end{table*}

\section{Reflection symmetry in band insulators}
\label{reflectionband}

To describe our schemes for reflection symmetric
topological insulators and superconductors,
in the following we will start from a tight-binding Hamiltonian.
We will focus on electronic insulators,
{\it i.e.,} systems with a conserved U(1) charge,
but a similar tight-binding formalism can be developed
for BdG Hamiltonians of topological superconductors.
Let us consider a tight-binding Hamiltonian,
\begin{align}
H
&=
\sum_{r,r'}
\psi^{\dag}(r)\,
\mathcal{H}(r,r')\,
\psi(r'),
\end{align}
where
$\psi(r)
$
is a $N_f$-component
fermion annihilation operator,
and index $r=(r_1,r_2,\ldots, r_d)$ labels a site
on a $d$-dimensional lattice
(the internal indices are suppressed).
Each block in the single particle Hamiltonian
$\mathcal{H}(r,r')$ is
an $N_f\times N_f$ matrix,
satisfying the hermiticity condition
$\mathcal{H}^{\dag}(r',r)
=\mathcal{H}(r,r')$,
and we assume the total size of the
single particle Hamiltonian is
$N_f V\times N_f V$,
where $V$ is the total number of lattice sites.
The components in $\psi(r)$
can describe, {\it e.g.},
orbitals or spin degrees of freedom,
as well as
different sites within a crystal unit cell
centered at $r$.

Provided the system has translational symmetry,
$\mathcal{H}(r,r') =
\mathcal{H}(r-r')$,
with periodic boundary conditions in each spatial direction
({\it i.e.}, the system is defined on a torus $T^d$),
we can perform the Fourier transformation and obtain
in momentum space
\begin{align}
H
=
\sum_{k \in \mathrm{BZ}}
\psi^{\dag}(k) \,
\mathcal{H}(k)\,
\psi (k) ,
\end{align}
where the crystal momentum $k$
runs over the first Brillouin zone (BZ),
and the Fourier component of the fermion operator
and the Hamiltonian are given by
$\psi(r)
=
V^{-1/2}
\sum_{k\in \mathrm{BZ}}
e^{{i} k \cdot r}
\psi(k)
$
and
$
\mathcal{H}(k)
=
\sum_{r}
e^{- {i} {k}\cdot {r}}
\mathcal{H}({r})
$,
respectively.
The Bloch Hamiltonian $\mathcal{H}(k)$
is diagonalized by
\begin{eqnarray}
\mathcal{H}(k)
|u^{{a}}({k}) \rangle
=
\varepsilon^{a} ({k})|u^{a}({k}) \rangle,
\quad
{a} = 1,\ldots, N_{f},
\end{eqnarray}
where
$|u^{{a}}({k}) \rangle$ is the
${a}$-th Bloch
wavefunction with energy
$\varepsilon^{{a}} ({k})$.
We assume that
there is a finite gap at the Fermi level,
and therefore
we obtain the unique ground state
by filling all states below
the Fermi level.
(In this manuscript, we always adjust
$\varepsilon^{a}({k})$ in such a way
that the Fermi level is at zero energy.)
We assume there are $N_-$ ($N_+$)
occupied (unoccupied) Bloch wavefunctions
with $N_{+} + N_{-} = N_{f}$.
We call the set of filled/unfilled Bloch wavefunctions
as
$
\{
|u^{-}_{\hat a}({k}) \rangle
\}
\equiv
\{
|v_{\hat a}({k}) \rangle
\}
$,
$
\{
|u^{+}_{a}({k}) \rangle
\}
\equiv
\{
|w_{a}({k}) \rangle
\}
$,
respectively,
where
hatted indices $\hat{a}=1,\ldots, N_-$
labels the occupied bands only.

In discussing symmetry protected topological phases,
we consider a set of (discrete) symmetry conditions
imposed on the tight-binding Hamiltonians.
Altland-Zirnbauer discrete symmetries,
{\it i.e.},
TRS, PHS, and chiral symmetry,
act, on the Bloch Hamiltonian, as
\begin{align}
T^{-1}\mH(-{k})T &=\mH({k}),
\nonumber \\
C^{-1}\mH(-{k})C &=-\mH({k}),
\nonumber \\
  S^{-1}\mH({k})S &=-\mH({k}),
  \label{chiral}
\end{align}
respectively,
where
$T$ and
$C$
are
antilinear operators,
and $S$ is a unitary operator.
These are on-site (purely local) symmetries.
While
PHS can most naturally be introduced
in the context of BdG Hamiltonians,
one can still impose a PHS
for electronic systems
with conserved particle number.

On the other hand,
reflection ($\mathcal{R}$) is a non-local operation;
by definition,
a reflection $\mathcal{R}$ in $x$-direction (= $r_1$-direction), say,
connects fermion operators
at $r=(r_1,r_2,\ldots, r_d)$
and at $\tilde{r}\equiv (-r_1,r_2,\ldots, r_d)$,
as
\begin{align}
\mathcal{R}\,
\psi(r) \,
\mathcal{R}^{-1}
=
R
\psi(\tilde{r}),
\end{align}
where $R$ is an $N_f\times N_f$ unitary matrix implementing reflection.
The invariance of $H$ under
$\mathcal{R}$ implies,
in momentum space,
\begin{align}
R^{-1} \mathcal{H}(k) R
= \mathcal{H}(\tilde{k}) \label{reflection}
\end{align}
where
$\tilde{k} = (-k_1, k_2, \ldots) = (-k_1, k_{\perp})$.


For example, for a spineless system,
possible realizations of these symmetries are
$R=\bI$, $T=\bI \Theta$,
and $C=\tau_x \Theta$,
where
$\tau_x$ is the first Pauli matrix acting on
the particle-hole grading (in the BdG Hamiltonian),
and $\Theta$ is the complex conjugate operator.
For a spin-one-half system,
$R=i\sigma_x$,
\cite{Weinberg:2005fk}
$T=i\sigma_y \Theta$, and $C=\tau_x \Theta$,
where $\sigma_i$ indicates Pauli matrices acting on spin degrees of freedom.
We will consider more realizations of these symmetries later.

Before discussing different realizations,
we here note that,
when there is a conserved U(1) charge,
there is a phase ambiguity in the definition
of the reflection operator;
\cite{Weinberg:2005fk}
when a Hamiltonian is invariant under
a reflection,
$\mathcal{R}:
\psi(r) \to
R
\psi(\tilde{r})$,
the system is also invariant under the reflection
followed by a U(1) gauge transformation,
$
\psi(r) \,
\to
e^{ {i}\phi} R
\psi(\tilde{r})$,
where $e^{ {i}\phi}$ is an arbitrary
phase factor.
The combined transformation,
$\mathcal{R}':
\psi(r) \to
R'
\psi(\tilde{r})$
with
$R'=e^{i \phi}R$,
is also qualified to be called reflection operation.
This redefinition changes, {\it e.g.},
the eigenvalues of the reflection transformations.

In this manuscript, we require $R$ to be \emph{hermitian}.
For example, in the spin-$1/2$ case,
we add an extra $3\pi/2$ phase factor in $R$ so that $R=\sigma_x$
anticommutes with $T$ and $C$.
With this convention,
we construct the classification tables
in terms of possible commutation and anticommutation relations of $R$
with the three non-spatial symmetry operations.
In this regard we can display the classification tables
in a well-organized manner.
While this is a matter of convention in classifying electrical
insulators,
this may not be so in classifying superconductors (BdG systems).
(It should be noted that for a AZ symmetry class
that can be interpreted as a BdG systems, it can also be realized
as an electrical system with some fine tuning).

One reason for this convention is that for different choices for
the phase of of reflection operator ({\it e.g.}, $R'=e^{i \phi}R$),
its algebraic relation with $T$ and $C$ is different.

Note that while $\mathcal{R}, \mathcal{C}, \mathcal{T}$, when acting on a fermion operator, may not commute due to the phase factor $e^{i \phi}R$, they always commute when acting on any fermion bilinears; Physically, all of the point group symmetry operators are expected to commute with three non-spatial symmetry operators:  TR, PH, and chiral symmetry operators. Although the requirement of the hermiticity of $R$ may not correspond to the real system, it simplifies the classification tables.

We will consider the topological classification when \emph{hermitian} $R$ commutes or anticommutes with $T$, $C$, and $S$.
For simplicity,
we define $R_S,\ R_T,$ and $R_C$ obeying
\be
SRS^{-1}=R_SR,\ TRT^{-1}=R_TR,\ CRC^{-1}=R_CR.
\ee
Hence, $R_X=\pm 1$ indicates the commutation or anticommutation relation between $R$ and the non-spatial symmetry operator $X$.
Furthermore, for the complex symmetry classes,
we define the symbol of the reflection symmetry operator $R_{R_S}$ to display the algebraic relation between $R$ and $S$. For four real symmetry classes that preserve TRS and PHS,
the symbol $R_{R_T,R_C}$ shows the similar property for $T$ and $C$ and provides the relation between $R$ and $S$, which is the combination of $T$ and $C$. 
For the four other real symmetry classes that have only one non-spatial symmetry (TRS or PHS),  the symbol of the reflection symmetry operator $R_{R_{T/C}}$ is defined to show the algebraic relation between $R$ and the non-spatial symmetry operator. 
In short, class AIII, AI, D, AII, and C, which preserve only one non-spatial symmetry, have two possible reflection symmetry operators $R_{-}$ and $R_{+}$. On the other hand, class BDI, DIII, CII, and CI, which preserve TRS and PHS, possess four possible reflection symmetry operators $R_{--},\ R_{++},\ R_{-+},$ and $R_{+-}$.  
To further simplify our notations, we define in a \emph{real} symmetry class the reflection symmetry operator ($R_+$ and $R_{++}$) that commutes all non-spatial symmetry operators as $\mathcal{R}^+$. Similarly, $\mathcal{R}^-$ indicates the reflection symmetry operator ($R_-$ and $R_{--}$) anticommuting with $T$ and(or) $C$ in a \emph{real} symmetry class.

\section{Topological classification of Dirac Hamiltonians
without reflection symmetry}
\label{DiracHam}

\begin{table}
\begin{center}
\begin{tabular}{|c|cccccccc|ccc|}
\hline
\multicolumn{12}{|c|}{
The original classification table} \\ \hline
   $\mbox{AZ class} \backslash d$  & 0 & 1 & 2 & 3 & 4 & 5 & 6 & 7 & T & C & S  \\
\hline\hline
  A & $\mathbb{Z}$ & 0 & $\mathbb{Z}$ & 0 & $\mathbb{Z}$ & 0 & $\mathbb{Z}$ & 0             & 0 & 0 & 0    \\
  AIII & 0 & $\mathbb{Z}$ & 0 & $\mathbb{Z}$ & 0 & $\mathbb{Z}$ & 0 & $\mathbb{Z}$          & 0 & 0 & 1    \\  \hline

  AI & $\mathbb{Z}$ & 0 & 0 & 0 & $2\mathbb{Z}$ & 0 & $\mathbb{Z}_2$ & $\mathbb{Z}_2$    & $+$ & 0 & 0     \\
  BDI & $\mathbb{Z}_2$ & $\mathbb{Z}$ & 0 & 0 & 0 & $2\mathbb{Z}$ & 0 & $\mathbb{Z}_2$     & $+$ & $+$ & 1    \\
  D & $\mathbb{Z}_2$ & $\mathbb{Z}_2$ & $\mathbb{Z}$ & 0 & 0 & 0 & $2\mathbb{Z}$ & 0     & 0 & $+$ & 0     \\
  DIII & 0 & $\mathbb{Z}_2$ & $\mathbb{Z}_2$ & $\mathbb{Z}$ & 0 & 0 & 0 & $2\mathbb{Z}$  & $-$ & $+$ & 1     \\
  AII & $2\mathbb{Z}$ & 0 & $\mathbb{Z}_2$ & $\mathbb{Z}_2$ & $\mathbb{Z}$ & 0 & 0 & 0   & $-$ & 0 & 0     \\
  CII & 0 & $2\mathbb{Z}$ & 0 & $\mathbb{Z}_2$ & $\mathbb{Z}_2$ & $\mathbb{Z}$ & 0 & 0   & $-$ & $-$ & 1     \\
  C & 0 & 0 & $2\mathbb{Z}$ & 0 & $\mathbb{Z}_2$ & $\mathbb{Z}_2$ & $\mathbb{Z}$ & 0     & 0 & $-$ & 0     \\
  CI & 0 & 0 & 0 & $2\mathbb{Z}$ & 0 & $\mathbb{Z}_2$ & $\mathbb{Z}_2$ & $\mathbb{Z}$    & $+$ & $-$ & 1
   \\
\hline
\end{tabular}
\caption{The original classification table of topological insulators and superconductors without reflection symmetry:\cite{Kitaev,Schnyder:2008gf} The first column represents the names of the ten symmetry classes, associated with the presence or absence of TR, PH, and Chiral symmetries in the last three columns. The number $0$ in the last three columns denotes the absence of the symmetry. The numbers $+1$ and $-1$ denote the presence of the symmetry and indicate the signs of the square TR operator and the square PH operator. }
\label{originaltable}
\end{center}
\end{table}

To capture the essential topological features
in an efficient manner,
we use the minimal Dirac Hamiltonian method.\cite{unpublish3rd}
This method simply considers
the minimal matrix form of Dirac Hamiltonians
in each AZ symmetry class and spatial dimension.
(See below for more details.)
When applied to topological insulators and superconductors
without spatial symmetries,
this method reproduces the periodic table of
topological insulators and superconductors
in the AZ symmetry classes
(See \Cref{originaltable}).
Such a Dirac Hamiltonian represents
a generic Hamiltonian
with the same topological features
when the spectra of the two systems can be continuously deformed from one to the other without closing the bulk band gap. Therefore, we will still use the method of minimal Dirac Hamiltonians to classify topological phases of reflection symmetric topological insulators and superconductors.

First, let us review the method of \emph{minimal Dirac Hamiltonians}.  The minimal Dirac Hamiltonian in $d$ spatial dimensions which respects the set of symmetries under consideration is written as
\be
\mH=m\gamma_0+k_1\gamma_1+\sum_{i\neq 1}^dk_i\gamma_i \label{Hgamma},
\ee
where $m$ is a constant and gamma matrices $\gamma_i$ obey the anticommutation relations
\be
\{\gamma_i,\gamma_j\}=2\delta_{ij}\bI, ~~i=0,1,\cdots,d. \label{anticommute}
\ee

In this manuscript,
we consider the classification of reflection symmetric systems with three local symmetries: TRS, PHS, and chiral symmetry.
That is, we seek the topological features of the Altland-Zirnbauer
(AZ) symmetry classes
\cite{PhysRevB.55.1142} with reflection symmetry.
When the Dirac Hamiltonian
preserves TRS, PHS, and chiral symmetry,
the gamma matrices satisfy
\begin{align}
  &
  [\gamma_0,T]=0,\ \{\gamma_{i\neq 0},T\}=0,
\label{Tgamma}
\\
&
\{\gamma_0,C\}=0,\ [\gamma_{i\neq 0},C]=0,
\label{Cgamma}
\\
&
\{\gamma_i,S\}=0,
\label{Sgamma}
\end{align}
respectively.
In addition,
when the system preserves reflection symmetry,
each term in the Dirac Hamiltonian obeys
\be
\{\gamma_1,R\}=0,\ [\gamma_{i\neq 1}, R]=0.
\label{Rgamma}
\ee

Consider a system in a real symmetry class, which possesses TRS or(and) PHS. If a term has the same anticommutation or commutation relation
of $\gamma_0$ in \Cref{Tgamma} or(and) \Cref{Cgamma} and anticommutes with the other gamma matrices in the Dirac Hamiltonian, we call this term as a {\it mass\/} term. Similarly, if a term behaves like $\gamma_{i\neq 0}$ in \Cref{Tgamma} or(and) \Cref{Cgamma}, this term is identified as a {\it kinetic\/} term. Furthermore, if a mass term anticommutes with $\gamma_0$, this is called as an \emph{extra} mass term. For a complex symmetry class, we also can define an extra mass term, which anticommutes with each gamma matrix in the Dirac Hamiltonian.

The presence of an extra mass term in the Dirac Hamiltonian
plays an essential role in distinguishing topological phases in a system.
We consider a system that preserves non-spatial symmetries above with/without other symmetries.
If any extra mass term does not preserve the
system's symmetries,
this term cannot be added to the Hamiltonian. 
When $m=0$, the bulk spectrum becomes gapless.
With negative energies filled, we identify $m=0$
as a quantum phase transition point,
where the gap between the empty and occupied band closes at
${k}=0$.
That is,
the phases with
positive and negative $m$ are topologically different.
On the contrary, an extra mass term preserving
the system's symmetries
can be added
to
the Hamiltonian as a perturbation.
We define this term as a {\it  symmetry preserving\/} extra mass term
(SPEMT). It is worth to mention that by definition in a real symmetry class an extra mass term is always a SPEMT.
When $m$ varies from $-\infty$ to $\infty$, there are no gap closing points.
Therefore, the system for any $m$ is always in the same phase,
which is topologically {\it trivial\/}.


\subsection{Topological invariant `$0$'}

For the topological
classification of the AZ symmetry classes
there are three different kinds of topological invariants:
$0,\ \bZ_2,\ \bZ$.
For a given set of symmetries and spatial dimension,
we write down
a Dirac Hamiltonian
of the minimum matrix dimension,
which is in the form of \Cref{Hgamma}.
If a SPEMT $(\frak{M})$
is allowed to be added
to the Hamiltonian,
the system is always in the trivial phase;
we can classify this
phase as topological invariant `$0$'.
For the other two cases
($\bZ_2$ and $\bZ$)
any SPEMT does not exist in the minimal model.
\cite{unpublish3rd}
Therefore, the system has at least two different phases by varying $m$ in \Cref{Hgamma}. To distinguish $\bZ_2$ and $\bZ$, we need to enlarge the Hamiltonian and then check the presence of a SPEMT.

\subsection{Topological invariant `$\bZ_2$'}

While enlarging the Dirac Hamiltonian,
we consider in the new system two minimal Dirac Hamiltonians which may have the same or opposite orientations.
That is, one is given by \cref{Hgamma} and the other is in the form of \cref{Hgamma}
with some $\gamma_i \rightarrow -\gamma_i$.
Moreover, each new gamma matrix in the enlarged Hamiltonian must anticommute with each other and keep the original symmetries. The expression of the enlarged Hamiltonian of the two minimal Dirac Hamiltonians can be written as
\be
\mH_2=\sum_i k_{n_i}\gamma_{n_i}\otimes \sigma_z +\sum_{{\rm remain}} k_{n_j}\gamma_{n_j}\otimes \bI. \label{doublesize}
\ee
The orientation of the second minimal Dirac Hamiltonian is determined by $\sigma_z$
and to simplify the expression of the equation, let $m=k_0$. The first summation is over arbitrary set of $\gamma_{n_i}$ ($n_i=0,1,2,...,d$) and the second summation is 
over $\gamma_{n_i}$'s that are not picked up by the first summation.  For the system with a $\bZ_2$ topological invariant, a SPEMT can always be added to the enlarged Hamiltonian in \Cref{doublesize} so the system is in the trivial phase. Therefore, the corresponding symmetries and spatial dimension restrict that the system can be in the only two different phases when the system is characterized by the sign of $m$ in the minimal Dirac Hamiltonian.

\subsection{Topological invariant `$\bZ(2\bZ)$'}
For the system with a $\bZ$ (or $2\bZ)$ topological invariant,
when the first summation in Eq. (\ref{doublesize}) includes {\it odd\/} number of $\gamma_{n_i}$'s, a SPEMT can be treated as a perturbation added into the Hamiltonian. However, when there are {\it even \/} number of $\gamma_{n_i}$'s in the first summation, a SPEMT does not exist. Therefore, the system can go through a quantum phase transition as $m$ varies from positive to negative.

To explain the $\bZ(2\bZ)$ invariant, we consider
$n$
copies of the minimal Dirac Hamiltonian but with different $m$'s
\be
\mH_m=\gamma_{n_i}\otimes
\begin{pmatrix}
m_1 & 0 & 0 & 0 \\
0 & m_2 & 0 &0 \\
0 & 0 & \ddots & 0 \\
0 & 0 & 0 & m_n
\end{pmatrix}
 +\sum_{i=1}^d k_{n_j}\gamma_{i}\otimes \bI. \label{ZbigH}
\ee
Assume all $m_i$'s are positive.
When one of $m_i$'s varies
from positive to negative, the system goes through a quantum phase transition.
This phase transition cannot be avoided
in the absence of SPEMTs due to the $\bZ(2\bZ)$ invariant\cite{unpublish3rd}.
By adjusting $m_i$'s the system passes through $n$ times of different quantum phase transitions.
Hence, in different $m_i$'s, $n+1$ different quantum phases, which is labeled by the $\bZ$ topological invariant, describe the system.

In a system possessing a $\bZ$ (or $2\bZ$) topological invariant,
we compute the topological number of the Dirac Hamiltonian in the form of \Cref{ZbigH}.
Let $m_i=M-k^2$.
By calculating the winding number
and the Chern number (see appendix \ref{Zcal}),
the topological number in this system is $n(2n)$
if $M$ is positive and $0$ if $M$ is negative.
As $M$ is positive,
one sign flipping
($\gamma_i\rightarrow -\gamma_i$) causes that the topological number changes
its sign.
When the two systems with and without the sign switching
are coupled
a SPEMT can be present in the mixed Hamiltonian.
The entire system in the trivial phase is consistent with the zero value of the topological number ($n-n=0$).


\subsection{The correspondence between gapless edge states and bulk topology}

The topologically protected gapless edge states are present on the boundary between the trivial and non-trivial phases.
The presence and absence of such gapless edge states
determines the topological bulk phases:
(i) the \emph{presence} of {\it intact\/} gapless edge states implies
the non-triviality in the bulk.
(ii) the \emph{absence} of such states guarantees the system in the trivial phase.
However, the latter statement is not always true\cite{Hughes:2011uq} when \emph{spatial} symmetries are introduced.
In the following, we will show that this statement holds when a system only preserves TRS, PHS, or chiral symmetry.
Furthermore, for reflection symmetry this correspondence between bulk and boundaries is also true,
when we choose the boundaries reflected to themselves and the translation symmetry in the reflection direction is preserved.

First, consider the minimal Dirac Hamiltonian in \Cref{Hgamma} in a symmetry class with $m=M-k^2$. Moreover, we consider a domain wall in the $r_d$ direction: let $M=M_0$ be a positive constant in the region $r_d>0$, which is in the non-trivial phase. For the trivial phase region $r_d<0$, let $M=-M_0$ be a negative constant. Therefore, $k_d$ is not a good quantum number. We replace $k_d$ by $-i\partial/\partial r_d$. The Dirac Hamiltonian can be rewritten as
\begin{align}
\mH=&\gamma_0(m\bI-i\gamma_0\gamma_d\frac{\partial}{\partial r_d})+\sum_{i=1}^{d-1}k_i\gamma_i,
\end{align}
where $m=M+(\frac{\partial}{\partial r_d})^2-\sum_{i=1}^{d-1}k_i^2$. To have the gapless energy states, we expect to find the wavefunctions so that the terms in the parenthesis vanish. To satisfy this vanishing condition, there are two possible solutions $i\gamma_0\gamma_d\vec{\phi}=\pm\vec{\phi}$. We choose the minus sign to have the normalizable wavefunctions
\ba
\Phi(r_d>0)&=&(c_1e^{-\frac{1}{2}(1-m_-)r_d}+c_2e^{-\frac{1}{2}(1+m_-)r_d})\vec{\phi},  \nonumber \\
\Phi(r_d<0)&=&e^{-\frac{1}{2}(1-m_+)r_d}\vec{\phi},
\ea
where $m_\pm=\sqrt{1\pm 4M_0-4\sum_{i=1}^{d-1}k_i^2}$. Our focus is on the low energy spectrum near $k=0$ so $M_0>\sum_{i=1}^{d-1}k_i^2$. Therefore, $m_+>1$ and $\mathrm{Re}(m_-)<1$ show that the wavefunction is normalizable.

Because $i\gamma_0\gamma_d$ commutes with $\gamma_{i\neq 0,\ d}$,
by using the
projection operator ${\bf P}=(\bI-i\gamma_0\gamma_d)/2$ we can discuss the projective Hamiltonian for the edge states in the ($i\gamma_0\gamma_d=-1$) eigenspace \be
\mH_{{\rm eff}}=\sum_{i=1}^{d-1}k_i\gamma_{{\bf p}i},
\ee
where $\gamma_{{\bf p}i}={\bf P}\gamma_i{\bf P}$. The energy spectrum ($\pm\sqrt{\sum_{i=1}^{d-1}k_i^2}$) shows the gapless behavior of the edge states.
Using the projective Hamiltonian, we can prove \emph{statement} (i) by considering a domain wall. When both sides of the domain wall are in the trivial phase, a SPEMT($\Gamma$) can be added into the Hamiltonian. Because $\Gamma$ anticommutes with all of the other gamma matrix, $\gamma={\bf P}\Gamma{\bf P}$ does not vanish and anticommutes with all of $\gamma_{{\bf p}i}$'s. Therefore, the gapless edge states become gapped. Furthermore, this statement implies that when the gapless edge states are intact, at least one side of the domain wall must be in a non-trivial phase.

To investigate \emph{statement} (ii), we focus on the behavior of the bulk Hamiltonian when the edge states are gapped without breaking any symmetry. The only one way to gap the {\it edge states} in $\mH_{{\rm eff}}$ is to add a symmetry preserving term that anticommutes with $H_{{\rm eff}}$, say $\tilde{\gamma}$. In the bulk Hamiltonian there exists corresponding symmetry preserving $\tilde{\Gamma}$ so that $\tilde{\gamma}={\bf P}\tilde{\Gamma}{\bf P}$.
Therefore, $\tilde{\Gamma}$ must commutes with $i\gamma_0\gamma_d$.\footnote{In the proper basis, $i\gamma_0\gamma_d$ can be written as $\sigma_z\otimes \bI$. Therefore, The decomposition of $\tilde{\Gamma}$ is uniquely in the form of $\tilde{\Gamma}_C+\tilde{\Gamma}_A$, where $[i\gamma_0\gamma_d,\tilde{\Gamma}_C]=0$ and $\{i\gamma_0\gamma_d,\tilde{\Gamma}_A\}=0$. Let $W$ be system's symmetry operator. We know $W\tilde{\Gamma}W^{-1}=\tilde{\Gamma}$ and $Wi\gamma_0\gamma_dW^{-1}=\pm i\gamma_0\gamma_d$. Hence, $[W\tilde{\Gamma}_{C}W^{-1},i\gamma_0\gamma_d]=0$ so $W\tilde{\Gamma}_{C}W^{-1}=\tilde{\Gamma}_{C}$. That is, $\tilde{\Gamma}_{C}$ preserves system's symmetry. Now let a new $\tilde{\Gamma}$ be $\tilde{\Gamma}_C$, which commutes with  $i\gamma_{0}\gamma_{d}$.
}
There are two possibilities of commutation and anticommutation relations of $\tilde{\Gamma}$ with each $\gamma_i$. First, $\tilde{\Gamma}$ anticommutes with each $\gamma_i$.\footnote{We know that $\tilde{\gamma}\gamma_{{\bf p} j}=-\gamma_{{\bf p} j}\tilde{\gamma}$, where $j\neq 0,\ d$. Therefore, ${\bf P}\tilde{\Gamma}{\bf P}^2\gamma_j{\bf P}=-{\bf P}\gamma_j{\bf P}^2\tilde{\Gamma}{\bf P}$. Since $\gamma_j$ and $\tilde{\Gamma}$ commute with $i\gamma_0\gamma_d$, ${\bf P}\tilde{\Gamma}\gamma_j=-{\bf P}\gamma_j\tilde{\Gamma}$. Similarly, consider the domain wall with $M\rightarrow -M$. The projection operator becomes ${\bf P}'=(\bI+i\gamma_0\gamma_d)/2$. We have ${\bf P}'\tilde{\Gamma}\gamma_j=-{\bf P}'\gamma_j\tilde{\Gamma}$. Thus, $\tilde{\Gamma}\gamma_j=-\gamma_j\tilde{\Gamma}$.
}
Second, $\tilde{\Gamma}$ anticommutes with $\gamma_{i\neq 0,d}$ but commutes with $\gamma_0$ and $\gamma_d$.\footnote{In the proper basis, $\gamma_0=\sigma_z\otimes \bI$. Therefore, $\tilde{\Gamma}$ can be written in the form of $\tilde{\Gamma}'_C+\tilde{\Gamma}'_A$, where $\tilde{\Gamma}'_C$ commutes with $\gamma_0$ and  $\tilde{\Gamma}'_A$ anticommutes with $\gamma_0$. Following the similar derivation with Ref.\ 37, $\tilde{\Gamma}'_C$ and $\tilde{\Gamma}'_A$ both preserve system's symmetry.}
In the first case $\tilde{\Gamma}$ plays a role of SPEMT keeping the system in the trivial phase. The second case needs to be investigated scrupulously. Although $\tilde{\Gamma}$ is not a SPEMT, in the case of some specific symmetries there exist a SPEMT, which is a hermitian matrix $i\tilde{\Gamma}\gamma_0\gamma_d$. If the system preserves TRS, PHS, and chiral symmetry, then by \Cref{Sgamma,Tgamma,Cgamma} $i\tilde{\Gamma}\gamma_0\gamma_d$ also preserves those symmetries. The correspondence between edge states and bulk topology can be applied for these three symmetries.
	
	Let reflection symmetry reflect only in the $k_d$ direction, then $\{R,\gamma_d\}=0$ and $[R,\gamma_{i\neq d}]=0$. Therefore, $i\tilde{\Gamma}\gamma_0\gamma_d$ breaks reflection symmetry and then the absence of the gapless edge states does not imply the trivial bulk topology. However, if reflection symmetry is not in the $k_d$ direction, then $i\tilde{\Gamma}\gamma_0\gamma_d$ preserves the symmetry and can be present in the Hamiltonian as a SPEMT. The gapped edge states possessing reflection symmetry guarantee the triviality in bulk.
	

	When the translational symmetry in the reflection direction is broken,
	the correspondence between gapless edge states and bulk topology does not hold. However, we still can use the mid-gap states in the entanglement spectrum to distinguish topological trivial and non-trivial phase.\cite{Hughes:2011uq,Turner:2010qf}
	We leave this issue in the future discussion. In the manuscript, we use the existence of SPEMTs in the minimal Dirac Hamiltonians to determine possible topological phases.

\section{The classification of $R_+,\ \mathcal{R}^+$ and $\mathcal{R}^-$ --  shifted periodic table}\label{shifttable}

We consider the \emph{real} AZ symmetry classes
with the reflection symmetry
commuting $(\mathcal{R}^+)$ and anticommuting $(\mathcal{R}^-)$
with $T$ and $C$.
For $\mathcal{R}^+$, the classification table is obtained from
the original table without reflection symmetry
by ``upward shift'' in spatial dimensions as shown
in Table \ref{commutationtable}.
The topological invariant $\bZ$ is replaced by
a new topological invariant $M\bZ$,
which will be explained later
(``mirror'' topological invariant).
Similarly, for $\mathcal{R}^-$, in $d$ dimensions
the topological invariants in the new table
is the ones in $d+1$ dimensions
in the original table
(``downward shift''),
except for the absence of the second descendant
$\bZ_2$\cite{SRFLnewJphys} of $\bZ$ as shown
in \Cref{anticommutationtable}.
In the following,
for the real symmetry classes,
we will construct these two tables
for
$\mathcal{R}^-$ and $\mathcal{R}^+$ and define
the topological invariant $M\bZ$.
We leave the discussion of the \emph{complex} symmetry
classes ($R_+$) for interested readers.

 \subsection{Classification of Dirac Hamiltonians
 }

Let us start by
giving a brief description of the mechanism
behind
these dimensional shifts.
By knowing
that $\gamma_1$ anticommutes with all of the other gamma matrices, we construct a hermitian matrix $i\gamma_1 R$ satisfying the anticommutation relation
\be
\{i\gamma_1R, \mH\}=0.
\ee
First,
for the case of $\mathcal{R}^-$,
from  \Cref{Rgamma,Tgamma,Cgamma} $i\gamma_1R$
can be used as a gamma matrix to construct another Dirac kinetic term
in one higher dimensions ($d+1$), say as $\gamma_{d+1}$.
Because
$[i\gamma_1R,C]=0$ and $\{i\gamma_1R,T\}=0$,
the $(d+1)$-dimensional Hamiltonian preserves the same set of
local AZ symmetries.
Alternatively,
to construct a Hamiltonian in $d$ dimensions with $\mathcal{R}^-$,
we can start from a system in $d+1$ spatial dimensions
preserving the same local symmetries, but not reflection.
By removing one gamma matrix $\gamma_{d+1}$ (and momentum component) from the kinetic,
we obtain the $d$-dimensional Hamiltonian with reflection symmetry.
We will later make use of the topological classification
 of the $d+1$-dimensional Hamiltonians without reflection
 to discuss the topological classification of our $d$-dimensional
 target Hamiltonians with reflection symmetry.

On the other hand, for the case of $\mathcal{R}^+$,
$i\gamma_1 R$ can be used as an extra mass term:
it can be added to the Hamiltonian
without changing its AZ symmetry class
since $\{i\gamma_1R,C\}=0$ and $[i\gamma_1R,T]=0$,
while $i\gamma_1R$ breaks the reflection symmetry.
Because of the algebraic structure of the Clifford algebra,
adding a mass term effectively acts as removing one kinetic gamma matrix,
and therefore, effectively decreases the spatial dimension by one.
\cite{unpublish3rd}
Therefore, the topological classification of the $d$-dimensional Hamiltonians
with $\mathcal{R}^+$ is related to the classification of $(d-1)$-dimensional Hamiltonians
in the corresponding AZ symmetry class without reflection symmetry.
This ``upward'' shift is also supported by considering
the Hamiltonian in the $(d-1)$-dimensional mirror plane in the
Brillouin zone.
The topological invariant defined for the $(d-1)$-dimensional Hamiltonian
directly determines the topological class of the original $d$-dimensional Hamiltonian
with reflection symmetry.
(See below for more details).

For those two cases ($\mathcal{R}^-$ and $\mathcal{R}^+$),
when a SPEMT in $d\pm 1$ dimensions in the AZ symmetry class without reflection symmetry is still a SPEMT
in $d$ dimensions with reflection symmetry,
both of the systems share the same topological invariant.
In the following we will show that when
 systems in $d\pm 1$ dimensions possesses `$0$' and $\bZ_2$
topological invariants, the corresponding reflectional systems have the same topological invariants. 
Likewise, for a $\bZ$ invariant in $d\pm 1$ dimensions, the corresponding reflection symmetric system in $d$ dimension has a $\bZ$-like topological invariant.

\paragraph{Topological invariant `0'}

Consider a system
in an AZ symmetry class in $d\pm 1$
that
has a `0' topological invariant.
Therefore, the presence of a SPEMT ($\frak{M}$) in the minimal Dirac Hamiltonian
in \Cref{Hgamma} keeps the system in the trivial phase.
We define the reflection operator $R=i\Lambda\gamma_1$,
where $\Lambda=\tilde{\gamma}_1(\gamma_{d+1})$ corresponds to $\mathcal{R}^+(\mathcal{R}^-)$.
Because $\mM$ anticommutes with $\Lambda$ and $\gamma_1$, the system in $d$ spatial dimensions with $\mM$
preserves reflection symmetry.
The same AZ symmetry class with reflection symmetry in $d$ dimensions has a `0' topological invariant.

\paragraph{Topological invariant `$\mathbb{Z}_2$'}

For a $\bZ_2$ topological invariant, in $d\pm 1$ dimensions the minimal Dirac Hamiltonian in \Cref{Hgamma} in one of the AZ symmetry classes has no SPEMTs. Therefore, SPEMTs do not exist for the minimal Dirac Hamiltonian in $d$ dimensions with reflection symmetry $R=i\Lambda\gamma_1$. Again to find the topological property for the reflection symmetry, the minimal Dirac Hamiltonian in $d$ dimensions can be enlarged in several ways:
\begin{align}
  \mH^d_{2}&=
  k_1\gamma_1\otimes \bI+\sum_{n_i\neq 1} k_{n_i}\gamma_{n_i}\otimes \sigma_z
  +\sum_{{\rm remain}} k_{n_j}\gamma_{n_j}\otimes \bI,
  \label{dbI} \\
  {\mH^d_{2}}'&=
  k_1\gamma_1\otimes \sigma_z+\sum_{n_i\neq 1} k_{n_i}\gamma_{n_i}\otimes \sigma_z
  +\sum_{{\rm remain}} k_{n_j}\gamma_{n_j}\otimes \bI,
  \label{dsigmaz}
\end{align}
with the unchanged reflection symmetry operator $R=i\Lambda\gamma_1\otimes \bI$. Because of the $\bZ_2$ topological invariant,
for
$\mathcal{H}^d_2$ we can construct the Hamiltonian in $d\pm 1$ dimensions in the same AZ symmetry class with a SPEMT $\mM$ without the reflection symmetry as
\be
\mathcal{H}^{d\pm 1}_2
=\mH^d_{2}+\lambda\Lambda\otimes \bI +\mM,
\label{d1bI}
\ee
where $\lambda=k_{d+1}(\tilde{m})$ in $d+1(d-1)$ dimensions. To check whether $\mM$ preserves the reflection symmetry, the commutation relation between $R=i\Lambda\gamma_1\otimes\bI$ and $\mM$ should be considered. By the definition of a SPEMT, $\mM$ anticommutes with $\Lambda\otimes \bI$ and $\gamma_1\otimes \bI$. Therefore, $\mM$ preserves the reflection symmetry. The Hamiltonian can be gapped out without breaking any symmetry.

For the second Hamiltonian ${\mH^d_{2}}'$, the corresponding gapped Hamiltonian in $d\pm 1$ dimensions is written as
\be
{\mathcal{H}^{d\pm 1}_2}'
={\mH^d_{2}}'+\lambda\Lambda\otimes \sigma_z +\mM. \label{d1sigmaz}
\ee
Because the SPEMT $\mM$ commutes with $\gamma_1\otimes \bI$ and $\Lambda\otimes\bI$, $\mM$ preserves the reflection symmetry.
This is so since
if one of $\gamma_1\otimes \bI$ and $\Lambda\otimes\bI$ anticommutes with $\mM$,  $\bI\otimes\sigma_z$ must commute with $\mM$ so
both of $\gamma_i\otimes \bI$ and $\Lambda\otimes\bI$
anticommute
with $\mM$.
This contradicts with the assumption that
there are no SPEMTs in the minimal Hamiltonian.
In short, the $d$-dimensional system with reflection symmetry
has a $\bZ_2$ invariant inherited
from the $d\pm 1$-dimensional system without reflection symmetry. However, there is an \emph{exception}. For the anticommutation case a reflection system from $d+1$ dimensions corresponding to the second descendants\cite{SRFLnewJphys} $\bZ_2$ of $\bZ$ has `$0$' topological invariant instead of $\bZ$, which will be discussed later.

\paragraph{Topological invariant `$\mathbb{Z}$'}

Consider a system in $d\pm 1$ dimensions in an AZ symmetry class
that
has a $\bZ$ topological invariant. Therefore, in the corresponding $d$-dimensional system with reflection symmetry any SPEMT does not exist for the minimal Dirac Hamiltonian, which is similar with the $\bZ_2$ case.
To distinguish the topological invariant from $\bZ_2$,
we need to enlarge the minimal Dirac Hamiltonian
in the forms of \Cref{dbI,dsigmaz}.
The corresponding Hamiltonians without reflection symmetry are written in the forms of \Cref{d1bI,d1sigmaz}
respectively so that a SPEMT $\mM$ preserves the reflection symmetry.
However, by the definition of the $\bZ$ topological invariant,
the SPEMT $\mM$ are present in \Cref{d1bI,d1sigmaz} only when the first summation is over {\it odd\/} number of
the gamma matrices.
Thus, for the system with reflection symmetry the presence/absence of a SPEMT is determined by the first summation odd/even number of gamma matrices in \Cref{dbI,dsigmaz} but does not depend on the way to enlarge $\gamma_1$.
In the next paragraph we will prove that such a system possesses a topological invariant.
We label this invariant by `$M\bZ$'.
The reason is that the $M\bZ$ system behaves the same with the $\bZ$ one in the non-reflectional symmetry direction but topological property of $M\bZ$ is insensitive in the reflectional symmetry direction. Moreover, we will show that $M\bZ$ number is defined in the reflection(mirror) symmetry planes ($k_1=0$ or $\pi$) so $M$ means mirror.

\begin{table}
\begin{center}
\begin{tabular}{|c|cccccc|ccc|}
\hline
\multicolumn{10}{|c|}{
Commutation relations of $R$} \\
\hline
   $\mbox{AZ class+R} \backslash d$   & 1 & 2 & 3 & 4 & 5 & 6  & T & C & S  \\
\hline\hline
  A & $M\bZ$ & 0 & $M\bZ$ & 0 & $M\bZ$ & 0      & 0 & 0 & 0    \\
  AIII & 0 & $M\bZ$ & 0 & $M\bZ$ & 0 & $M\bZ$ & 0 & 0 & 1    \\  \hline
  AI & $M\bZ$ & 0 & 0 & 0 & $2M\bZ$ & 0   & $+$ & 0 & 0     \\
  BDI & $\mathbb{Z}_2$ & $M\bZ$ & 0 & 0 & 0 & $2M\bZ$      & $+$ & $+$ & 1    \\
  D & $\mathbb{Z}_2$ & $\mathbb{Z}_2$ & $M\bZ$ & 0 & 0 & 0     & 0 & $+$ & 0     \\
  DIII & 0 & $\mathbb{Z}_2$ & $\mathbb{Z}_2$ & $M\bZ$ & 0 & 0  & $-$ & $+$ & 1     \\
  AII & $2M\bZ$ & 0 & $\mathbb{Z}_2$ & $\mathbb{Z}_2$ & $M\bZ$ & 0  & $-$ & 0 & 0     \\
  CII & 0 & $2M\bZ$ & 0 & $\mathbb{Z}_2$ & $\mathbb{Z}_2$ & $M\bZ$   & $-$ & $-$ & 1     \\
  C & 0 & 0 & $2M\bZ$ & 0 & $\mathbb{Z}_2$ & $\mathbb{Z}_2$    & 0 & $-$ & 0     \\
  CI & 0 & 0 & 0 & $2M\bZ$ & 0 & $\mathbb{Z}_2$    & $+$ & $-$ & 1
   \\
\hline
\end{tabular}
\caption{
  \label{commutationtable}
 The classification table for $R$(class A), $R_+$(class AIII), and  $\mathcal{R}^+$(real symmetry classes).
Each non-spatial symmetry operator commutes with $R$. For class A, a system has only reflection symmetry so no commutation issue is in this class.  If we treat $M\bZ(2M\bZ)$ as $\bZ(2\bZ)$, this table is obtained from the original table just by ``upward shift'' in spatial dimensions. }
\end{center}
\end{table}

\begin{table}
\begin{center}
\begin{tabular}{|c|cccccc|ccc|}
\hline
\multicolumn{10}{|c|}{
Anticommutation relations of $R$} \\
\hline
   $\mbox{AZ class+R} \backslash d$   & 1 & 2 & 3 & 4 & 5 & 6  & T & C & S  \\
\hline\hline
  AI   & 0 & 0 & $2M\bZ$ & 0 & 0 & $\mathbb{Z}_2$   & $+$ & 0 & 0     \\
  BDI  & 0 & 0 & 0 & $2M\bZ$ & 0 & 0     & $+$ & $+$ & 1    \\
  D  & $M\bZ$ & 0 & 0 & 0 & $2M\bZ$ & 0     & 0 & $+$ & 0     \\
  DIII  & $\mathbb{Z}_2$ & $M\bZ$ & 0 & 0 & 0 & $2M\bZ$ &  $-$ & $+$ & 1     \\
  AII  & 0 & $\mathbb{Z}_2$ & $M\bZ$ & 0 & 0 & 0  &  $-$ & 0 & 0     \\
  CII  & 0 & 0 & $\mathbb{Z}_2$ & $M\bZ$ & 0 & 0  & $-$ & $-$ & 1     \\
  C  & $2M\bZ$ & 0 & 0 & $\mathbb{Z}_2$ & $M\bZ$ & 0    & 0 & $-$ & 0     \\
  CI  & 0 & $2M\bZ$ & 0 & 0 & $\mathbb{Z}_2$ & $M\bZ$   & $+$ & $-$ & 1
   \\
\hline
\end{tabular}
\caption{
  \label{anticommutationtable}
 The classification table for $\mathcal{R}^-$.
For the eight real symmetry classes, $R$ anticommutes with $T$ and $C$ but commutes with $S$. For class AIII, the only non-spatial symmetry operator $S$ anticommutes with $R$. We treat $M\bZ(2M\bZ)$ as $\bZ(2\bZ)$, then the topological invariants in this table in $d$ dimensions is the ones in the original table in $d+1$ dimensions, except for the second descendant $\bZ_2$ of $\bZ$.}
\end{center}
\end{table}

\subsection{Topological numbers}

\subsubsection{Topological numbers of $M\bZ$}

For the case where
the reflection operator $R$ commutes with the non-spatial discrete symmetries,
the topological numbers $M\bZ$
can be defined from the bulk Hamiltonian to characterize bulk topology and protected gapless edge states.
The Hamiltonian without $k_1$ commutes with $R$;
therefore, the Hamiltonian can be
block-diagonalized in the two eigenspaces $R=\pm1$ because $R$ is hermitian.
Each individual block Hamiltonian is (not) invariant under the original symmetries
if such symmetry operators (anti)commute with $R$.
However, the Hamiltonian still belongs to one of the AZ symmetry classes, which corresponds to the non-spatial symmetry operators commuting with $R$.
We name this symmetry class in the mirror planes of 
$k_1=0$ or $\pi$ as a \emph{mirror symmetry class}. The details of mirror symmetry classes will be discussed in appendix \ref{mirror class}.

We focus on one of the blocks with a definite eigenvalue, $R=1$, say,
because the topological numbers of
these two blocks
differ by signs when the weak topological index vanishes.
The reason is that a $d$-dimensional system with a nonzero
weak index can be understood by a stacking limit of $d-1$ dimensional topologically nontrivial layers\cite{weakindices}. Furthermore, for any $k_1$ the sum of these two topological numbers is invariant;
hence, if this total number does not vanish, by definition the weak index is nonzero.
In the following, we always consider the case that the weak index vanishes to define the $M\bZ$ number.

\paragraph{$M\bZ$ for $\mathcal{R}^+$}

First,
suppose the non-spatial symmetry operators commute with $R$.
For a $d$-dimensional system possessing a $M\bZ$ topological invariant,
the topological number does not depend on the $k_1$
direction (direction of the reflection symmetry).
Furthermore, in $d-1$ dimensions the mirror symmetry class,
which is the same with the original AZ symmetry class,
has a $\bZ$ topological invariant.
Hence, to obtain the $M\bZ$ number,
we can calculate the $\bZ$ number in
one of the blocks with a definite reflection eigenvalue
in $d-1$ dimensions by using \Cref{oddZ} or \Cref{evenZ}
because the Dirac Hamiltonian without $\gamma_1$ commutes with $R$.
This $\bZ$ property is protected
by the corresponding block diagonal non-spatial symmetry operators
since $R$ commutes with these operators.

In the
continuum
model the $M\bZ$ number can be properly defined
for the block Hamiltonians at $k_1=0$.
However, in the lattice model,
which can be obtained by the replacement
  $k_1\rightarrow \sin n k_1$ ($n \in \mathbb{Z}$),
  $\sin n k_1$ vanishes in the Hamiltonian
only when $k_1=0, \pm \pi/n, \pm 2\pi/n, \ldots, \pm\pi$.
These points are the possible positions to have
a $(d-1)$-dimensional $\bZ$ number.
However, the $\bZ$ numbers
which are not at the symmetry points $(k_1= 0,\ \pi)$,
are fragile, or not protected:
They can vanish by coupling the opposite $\bZ$ numbers
in the other block of $R=-1$
without breaking reflection symmetry.
Furthermore, no SPEMTs are allowed in the bulk Dirac Hamiltonian around the symmetry points so the $\bZ$ numbers at the symmetry points are invariant. Therefore, we can calculate the two numbers $\nu_0^{d-1}$ and $\nu_\pi^{d-1}$ at $k_1=0,\ \pi$ respectively in the block of $R=1$.

To have the topological number of the strong index,
we need to consider translational symmetry breaking.
The presence of translational symmetry breaking along the $r_1$
direction
connects the $\bZ$ numbers at
the two symmetric points so the total topological invariant number
is $\nu_0^{d-1}+\nu_\pi^{d-1}$.
However, this number is not the strong index.
The strong index $N_{M\bZ}$ is given by
\be
N_{M\bZ}=\nu_0^{d-1}+\nu_\pi^{d-1}-2N_{\rm{weak}},
\ee
where $N_{\rm{weak}}$ is the {\it mirror} weak index,
which is the weak index in one of the blocks of $R=\pm 1$ and invariant
for any $k_1$.
Such a mirror weak index is determined by $d-1$-dimensional nontrivial
layers\cite{weakindices},
which are stacked to a nontrivial weak system.
To have the strong index, we determine the mirror weak index first by considering two possible situations:
$\nu^{d-1}_0\nu_\pi^{d-1}>0$ and $\nu^{d-1}_0\nu_\pi^{d-1}<0$.
On the one hand ($\nu^{d-1}_0\nu_\pi^{d-1}>0$), the mirror weak index is
\begin{align}
  N_{{\rm weak}}={\rm sgn}(\nu^{d-1}_0)
  {\rm min}(|\nu^{d-1}_0|,|\nu_\pi^{d-1}|).
  \label{mirror weak}
\end{align}
Because the total invariant number are the sum of $2N_{{\rm weak}}$ and the strong index, we can write the strong index topological number is 	
\be
{\rm sgn}(\nu^{d-1}_0)|\nu^{d-1}_0-\nu_\pi^{d-1}|.
\ee	
On the other hand, when $\nu^{d-1}_0\nu_\pi^{d-1}<0$ the mirror weak index is absent. The strong index is the total invariant number $\nu_0^{d-1}+\nu_\pi^{d-1}$. From these two cases, the $M\bZ$ number is defined as 	
\begin{align}
  N_{M\bZ}={\rm sgn}(\nu_0^{d-1}-\nu_\pi^{d-1})(|\nu_0^{d-1}|-|\nu_\pi^{d-1}|).
  \label{numSZ}
\end{align}
The signs determine the orientation; however, the $N_{M\bZ}$ does not have the summation property like $\bZ$ topological invariant ($N_\bZ$). Consider a system is the collection of several subsystems. Each subsystem has its own $\bZ$ number $N_{\bZ}^i$. Therefore, the $\bZ$ number of the entire system is given by
$
N_{\bZ}=\sum_iN_{\bZ}^i
$.
This relations does not hold for $M\bZ$.
To obtain $N_{M\bZ}$, we have to compute the two $\bZ$ numbers ($\nu_0^{d-1}$ and $\nu_\pi^{d-1}$) for the entire system in the reflection symmetric planes and then use \cref{numSZ}.

\paragraph{$M\bZ$ for $\mathcal{R}^-$}

Similarly, we consider the case that a $d$-dimensional system with reflection symmetry operator $\mathcal{R}^-$,
which anticommutes with the TRS operator or PHS operator.
If $d$ is even, the system possessing $M\bZ$ preserves Chiral symmetry.
This Chiral symmetry operator commutes with $\mathcal{R}^-$
so in the Hamiltonian block-diagonalized by $\mathcal{R}^-$ each block has the corresponding Chiral symmetry operator. Neither the TRS nor PHS is block-diagonalized at the same time because of the anticommutation relations.
Therefore, each block in the Hamiltonian belongs to class AIII in even dimensions and then the topological number can be evaluated as the winding number by \Cref{oddZ}. On the other band, if $d$ is odd, the off-diagonal Hamiltonians, which do not preserve any symmetry, belong to class A in odd dimensions. Hence, in class A the Chern number in \Cref{evenZ} can characterize the topological number at the symmetry points. In short, the topological number in the anticommutation case still can be described by \Cref{numSZ}.

\subsubsection{Topological numbers of $\bZ_2$}\label{bz2sec}

\paragraph{$\bZ_2$ for $\mathcal{R}^+$}

 Consider a $d$-dimensional $\bZ_2$ system with
 $\mathcal{R}^+$ commuting with the local symmetries.
In such a system, the topological invariant for
a block in the $R$-block-diagonal Hamiltonian is $\bZ_2$ in $d-1$
dimensions in the mirror symmetry class, which is the same with system's symmetry class.
Such a $\bZ_2$ topological invariant was already evaluated in several ways\cite{Kane:2005kx,Fu:2007uq,JPSJ.80.013602,Kitaev1D}.
 Therefore, at the two symmetry points $k_1=0,\ \pi$, the $\bZ_2$ numbers are defined as $\nu_0$ and $\nu_\pi$ respectively in the block of the Hamiltonian. The $\bZ_2$ number for the entire system is
 \be
 N_{\bZ_2}=\nu_0+\nu_\pi\ {\rm mod} \ 2. \label{Z2def}
 \ee
The reason is that the reflection symmetry does not prevent a translational symmetry breaking density
wave from coupling and gapping out a pair of bulk band gap closing (quantum phase transitions) at these
two symmetry points. Only one bulk band gap closing survives
under arbitrary symmetry preserving perturbations when a system possesses an odd number of closing.

 \paragraph{$\bZ_2$ for $\mathcal{R}^-$}
Consider in the case that reflection symmetry operator anticommutes with $T$ and(or) $C$, which is more complicated. We discuss two possible cases respectively: the first and second descendants $\bZ_2$ of $\bZ$ in $d+1$ dimensions.

First, consider a system with reflection symmetry corresponding to the first descendant $\bZ_2$ of $\bZ$ in $d+1$ dimensions.
We note that in $d$ dimensions
the original topological classification gives a $\bZ_2$ topological invariant. The topological number in this case can be defined by the original $\bZ_2$ number. That is, a system in such a symmetry class with and without the reflection symmetry has the same topological invariant.


Secondly, in a system with reflection symmetry corresponding to the second descendant $\bZ_2$ of $\bZ$ in $d+1$ dimensions, the topological number cannot be properly defined. The reason is that $T$ and(or) $C$ anticommute with $R$ so the mirror symmetry class is class A. Therefore, no $\bZ_2$ topological numbers can be defined at $k_1=0,\ \pi$. Furthermore, in $d$ dimensions the corresponding topological invariant is `$0$' in the original classification. It turns out that a SPEMT can be present in the Hamiltonian to prevent the bulk gap closing so this case is classified as `$0$'. The further discussion is in the following. Without enlarging the minimal Dirac Hamiltonian in $d$ dimensions in the corresponding symmetry class, $d+3$ kinetic gamma matrices $(\gamma_1,\gamma_2,\ldots,\gamma_{d+3})$ and one mass gamma matrix ($\gamma_0$)
can be present.
\cite{unpublish3rd,gamma_dplus3}
The reflection symmetry operator is defined as $R=i\gamma_0\gamma_{d+1}$ to satisfy the anticommutation relations with TRS and PHS operators. Due to the presence of $\gamma_{d+1},\ \gamma_{d+2},\ \gamma_{d+3}$, we have more choices to add some symmetry preserving terms in the minimal Dirac Hamiltonian in \Cref{Hgamma}
\be
\mH_{\delta}=m\gamma_0+k_1\gamma_1+\sum_{i\neq 1}^dk_i\gamma_i+\delta \Delta,	\label{deltashift}
\ee
where $\delta$ is a positive constant and $\Delta=i\gamma_1\gamma_{d+1}\gamma_{d+2}$, which is invariant under all system's symmetries. Since $\Delta$ commutes with only $\gamma_1$ in $\mH_{\delta}$, the eigenvalues of $\gamma_1(k_1\bI+\delta i\gamma_{d+1}\gamma_{d+2})$ are $k_1\pm\delta$ and $-k_1\pm\delta$  due to the eigenvalues $\pm 1$ of $i\gamma_{d+1}\gamma_{d+2}$. Therefore, when the quantum phase transition ($m=0$) occurs, the bulk gap closing points shift $k_1=\pm \delta$ and $k_\perp=0$.
Now we can add another symmetry preserving term to prevent the bulk gap closing at the new transition points by breaking translational symmetry.
This gap opening term is written in the form of the second quantization
\begin{align}
\hat{\frak{N}}=&\sum_{-\eta\leq k_1 < \eta}(ic_{k_1+\eta +\delta}^\dagger\frak{N} c_{k_1-\eta+\delta}+h.c.) \\ \nonumber
&+\sum_{-\eta< k_1 \leq \eta}(ic_{k_1+\eta -\delta}^\dagger\frak{N} c_{k_1-\eta-\delta}+h.c.), \label{NDW}
\end{align}
where $\eta$ is a positive constant less than $\delta$ and $\mN=i\gamma_{d+1}\gamma_{d+2}\gamma_{d+3}$, which anticommutes all of the terms in $\mH_{\delta}$. Also, $\hat{\mN}$ preserves TSR and PHS by \Cref{Tgamma,Cgamma}. By the definition of the reflection symmetry operator $\hat{R}=\sum_{k_1}c^\dagger_{k_1}(i\gamma_0\gamma_{d+1})c_{-k_1}$\cite{other:reflection},
it is easy to check that $\hat{\mN}$ preserves the reflection symmetry. The last thing we need to verify is that $\hat{\mN}$ prevents the bulk gap closing. To have the low energy spectrum, consider that case $m=0$ and $k_{\perp}=0$ so $\mH_\delta$ is a function of $k_1$. The Hamiltonian $\mH_\delta(k_1)$ with $c\hat{\mN}$ is in the form of the second quantization is written as
\begin{align}
\hat{\mH}_{\delta,c}
&=\sum_{-\eta \leq k_1 <\eta}(\Psi_{k_1+\delta}^\dagger
\mH_\eta(k_1+\delta)
\Psi_{k_1+\delta}  \nonumber \\
&+\Psi_{-k_1-\delta}^\dagger
\mH_\eta(-k_1-\delta)
\Psi_{-k_1-\delta}
) \nonumber \\
&+{\rm high\ energy\ terms},
\end{align}
where $\Psi_{p_1}=
\begin{pmatrix}
c_{p_1+\eta} & c_{p_1-\eta}
\end{pmatrix}^T$ and
\be
\mH_\eta(p_1)=\begin{pmatrix}
{\mH_\delta}(p_1+\eta) & ic\mN \\
-ic\mN & {\mH_\delta}'(p_1 -\eta) \\
\end{pmatrix}.
\ee
We compute the eigenvalues of the two blocks ($\mH_\eta(k_1+\delta)$ and $\mH_\eta(-k_1-\delta)$) to capture the low energy spectrum, because those two blocks are the reflection symmetry partners sharing the same energy spectrum. Therefore, consider the energy spectrum of one of the blocks, say
\begin{align}
\mH_\eta(k_1+\delta)&=(k_1+\delta)\bI\otimes \gamma_1+\eta \sigma_z\otimes \gamma_1 \nonumber \\
 &+\delta \bI \otimes \Delta +c \sigma_y \otimes \mN.
\end{align}
We note that $\mN$ anticommutes with $\Delta$ and $\gamma_1$ and $\Delta$ commutes with $\gamma_1$. Therefore, the expression of the energy square is
\be
E^2=(\eta\pm \sqrt{k_1^2+c^2})^2,\ (\eta\pm \sqrt{(k_1+2\delta )^2+c^2})^2. \label{gapDW}
\ee
Hence, when $c$ is larger than $\eta$, the energy never becomes zero. The bulk gap closing has been blocked by the symmetry preserving terms as SPEMTs. Therefore, the topological invariant in this case is `$0$'. In sec.\ \ref{egfakeZ2}, one example is provided to show that the translational symmetry breaking $\hat{\mN}$ gaps the edge states and destroys the mid-gap states in the entanglement spectrum.

\section{The classification of $R_{-+}$ and $R_{+-}$}\label{restclassification}

In this section, we consider
the topological classification
of insulators and superconductors
for the cases of $R_{-+}$ and $R_{+-}$.
To have these anticommutation and commutation relations
we have to consider the AZ symmetry classes that
preserve both TRS and PHS ---
class DIII, CII, CI, and BDI.
The results are summarized in \Cref{gamma1S,gammaleft}.

 \subsection{Classification of Dirac Hamiltonians with $R=i\gamma_1 S$
 }


As a start, let us consider, as a possible definition of reflection
operator, $R=i\gamma_1S$.
The commutation/anticommutation relations of
$R=i\gamma_1S$ with TRS and PHS operators are summarized in
in \Cref{gamma1S}.
They can be verified as follows.
Let us
go back to the expressions of TRS and PHS operators
\be
T=U_T\Theta,\ C=U_C\Theta,
\ee
where $U_T$ and $U_C$ are complex matrices. To simplify our problem, we assume that $U_T$ and $U_C$ are hermitian and unitary.
To define chiral operator $S$, which is hermitian,
we let $S=TC$ if $[U^*_C,U_T]=0$ or $S=iTC$ if $\{ U^*_C,U_T\}=0$. Therefore, $R=i\gamma_1S$ is hermitian. To determine the commutation and anticommutation relations of $R$ with $T$ and $C$, we have to check the relations of $S$ with $T$ and $C$. In the both cases ($[U^*_C,U_T]=0$ and $\{U^*_C,U_T\}=0$), we have the same relations
\begin{align}
TST^{-1}&=\pm S,
\quad
CSC^{-1}=\pm S, \label{TCS}
\end{align}
  where
  we pick up the plus sign in front of $S$
  when $T^2=\pm 1$ and $C^2=\pm 1$,
  whereas
  we pick up the minus sign
  when $T^2=\pm 1$ and $C^2=\mp 1$.
The reason is that
\begin{align}
\pm 1=T^2=U_TU_T^*,\ \pm 1=C^2=U_CU_C^*.
\end{align}
By using hermitian and unitary properties of $U_T$ and $U_C$,
\be
U_T=\pm U_T^*,\
U_C=\pm U_C^*.
\ee
By \Cref{Tgamma,Cgamma}, we obtain the relations exactly shown in \Cref{gamma1S}: $[T,R]=0$ and $\{C,R\}=0$ when $T^2=\pm 1$ and $C^2=\pm 1$ and $\{T,R\}=0$ and $[C,R]=0$ when $T^2=\pm 1$ and $C^2=\mp 1$.

	This construction of $R$ does not require any new gamma matrix. Hence, the absence of a SPEMT in the original classification is unchanged when the reflection symmetry is considered. When a SPEMT exists in the original Hamiltonian, we have to check whether this SPEMT preserves the reflection symmetry. If so, the system is in the trivial phase. If not, the reflection symmetry provides new topological phases.
We consider the following three cases separately.

\begin{table}
\begin{center}
\begin{tabular}{ |c | | c | c | c | c|| c|c|c|c|c|}
\hline			
 Class &   $TR$ &  $PH$& $Ch$  &  $R$   & $d=2$ & $d=3$ & $d=4$ & $d=5$ & $d=6$  \\
\hline
\hline
AIII & $0$ & $0$ & $1$ & $R_-$ & $0$ & $\bZ^1$ & $0$ & $\bZ^1$ & $0$ \\   
\hline
\hline	
  BDI &   $+1$  & $+1$ & $1$   		& $R_{+-}$ &   $0$ & 0 & 0 & $2\bZ^1$ & 0  \\
\hline
 DIII &  $-1$ & $+1 $& $1$ 	  		&  $R_{-+}$ &   $\bZ_2$ & $\bZ^1$ & 0 & 0 & 0  \\
\hline
 CII &   $-1$ & $-1$ & $1$ 	  		&  $R_{+-}$ &   0 & $\bZ_2$ & $\bZ_2$ & $\bZ^1$ & 0  \\
\hline
CI &  $+1$  & $-1$ & $1$ 					 &  $R_{-+}$ &   $0$ & $2\bZ^1$ & 0 & $\bZ_2$ & $\bZ_2$  \\
\hline
\end{tabular}
\caption{
The classification table for the case of reflection symmetry operator given by
$i\gamma_1S$.
By using the similar discussion of $R=i\gamma_1S$, a reflection symmetric system in class AIII $+R_{-}$ possesses a $\bZ^1$ invariant in odd dimensions and `$0$' invariant in even dimensions.
The AZ symmetry classes with the reflection symmetry have $\bZ_2$ and $\bZ^1$ corresponding to $\bZ_2$ and $\bZ$ without the reflection symmetry respectively. }
\label{gamma1S}
\end{center}
\end{table}

\begin{itemize}
\item
Let us first suppose
a system in an AZ symmetry class without reflection symmetry has `$0$' topological invariant.
A SPEMT ($\frak{M}$) exists in the minimal Hamiltonian. That is, $\frak{M}$ anticommutes with $S$ and $\gamma_i$. Therefore, when the reflection symmetry is imposed on the system, $R=i\gamma_1S$ commutes with $\frak{M}$ so the reflection symmetry is preserved. The system is still in the trivial phase and classified by the `$0$' topological invariant.

\item
	For
	the AZ symmetry class that has a $\bZ_2$ topological invariant,
no SPEMTs are allowed in the minimal Hamiltonian.
However, when the size of the Hamiltonian is doubled
in the form of \Cref{doublesize},
without considering reflection symmetry a SPEMT ($\frak{M}$) does exist.
We found that $\gamma_{n_i}\otimes \bI$
in \Cref{doublesize}
must commutes with $\frak{M}$
because if not, the minimal Hamiltonian can have a SPEMT.
Also, we know $\{\gamma_{n_j}\otimes\bI,\frak{M}\}=0$.
Therefore, if $n_i\neq 1$,
$[R=i\gamma_1S,\frak{M}]=0$ shows that the reflection symmetry is preserved.
Otherwise, the reflection symmetry is broken.
Hence, any SPEMT seems to be absent in this case;
however, later we will show that a SPEMT can be present,
which is similar to the counterfeit $\bZ_2$ case in sec.\ \ref{bz2sec}.
Hence, the reflection symmetry does not provide any extra topological
phase.
Such a system with the reflection symmetry is still classified as $\bZ_2$.


\item
Finally,
a system with a $\bZ$ topological invariant guarantees
no SPEMTs in the minimal Hamiltonian.
After doubling the size of the minimal Hamiltonian,
the original $\bZ$ topological invariant forbids
any SPEMT in \Cref{doublesize} when there are even terms
in the first summation.
When reflection symmetry is considered,
by the similar reasoning as the previous discussion,
$\{R=i\gamma_1 S,\frak{M}\}=0$
breaks reflection symmetry as $n_i=1$ in \Cref{doublesize}.
In short, the system can be gapped out by a SPEMT
only when the first summation in \Cref{doublesize}
is over an odd number of the gamma matrices, excluding $\gamma_1$.
Let us go back to the discussion of the $M\bZ$
topological invariant.
\Cref{dbI,dsigmaz} provide all of the possible twice-as-big minimal Dirac Hamiltonian. Only when  the first summation of \Cref{dbI} is over an odd number of the gamma matrices, a SPEMT can be present in the Hamiltonian.
Therefore, for an even number of the gamma matrices in the first summation of \Cref{dbI} a SPMET is forbidden by the $\bZ$ topological invariant.
In \Cref{dsigmaz} since $k_1\gamma_1\otimes \siz$ in the Hamiltonian with an even number of gamma matrices in the first summation,
a mass term preserving the non-spatial symmetries breaks the reflection symmetry.
Therefore, the bulk topology is non-trivial for both of the (the even numbers of gamma matrices). For the former,
the $\bZ$ number is non-zero as well as for the latter, the $M\bZ$ number does not vanishes.
Thus, the bulk topology \emph{might} be protected by $\bZ$ and $M\bZ$ topological invariants.
In the next subsection, we will show that for such a system the topological invariant labeled by $\bZ^1$ is determined by $\bZ$ and $M\bZ$ topological numbers.
\end{itemize}

  The above considerations lead to the classification summarized
  in Table \ref{gamma1S}. In the following,
we present a proper definition of topological invariants.

\subsubsection{Topological number of $\bZ^1$}{\label{bZ1}}
To define the proper $\bZ^1$ number,
characterizing bulk topology and intact gapless edge modes,
we consider $\bZ$ number ($N_\bZ$) and $M\bZ$ number ($N_{M\bZ}$)
as our candidates. In the following, we will prove that the $\bZ^1$ number is
\be
N_{\bZ^1}={\rm Max}(|N_\bZ|, |N_{M\bZ}|).
\label{NbZ1}
\ee

To simplify the problem, we consider the presence of non-trivial topology only at $k_1=0$ plane. In other words, $\bZ$ and $M\bZ$ numbers can be determined by the Hamiltonian around this symmetry plane.
We leave the general proof in appendix \ref{bZ1proof} for interested readers.
In a $d$-dimensional system the $\bZ$ number $N_\bZ$ and the $M\bZ$ number $N_{M\bZ}$, which is the $\bZ$ number $\nu_{d-1}$ at $k_1=0$, are both defined
by
either \Cref{oddZ} or \Cref{evenZ}.
We define $(\pm,\pm)$ and $(\pm,\mp)$ to describe the signs of the gamma matrices in \Cref{Hgamma}. The first slot of the parentheses indicates the sign of $\gamma_1$ and the second slot $+(-)$ shows even(odd) number of the other gamma matrices having the minus sign. According to the $\bZ^1$ properties, the system with $(+,+)$ and $(+,-)$ can be gapped out by a SPEMT. Similarly, the pair of $(-,-)$, $(-,+)$ are in the trivial phase, too. However, two minimal Hamiltonians from these two different pairs are protected by the $\bZ^1$ topological invariant. We can treat these two pairs be two independent systems of $(+,\pm)$ and $(-,\pm)$. Let the numbers of the minimal Hamiltonians of $(\pm,\pm)$ and $(\pm,\mp)$ be $N_{\pm,\pm}$ and $N_{\pm,\mp}$ respectively. A nontrivial minimal Hamiltonian provides one protected gapless edge mode so the numbers of intact gapless edge modes for these systems are given by respectively
\be
N_{+,\pm}=N_{+,+}-N_{+,-},\ N_{-,\pm}=N_{-,+}-N_{-,-}.
\ee
We note that a system $(+,\pm)$ has $N_\bZ=\pm 1$ and $N_{M\bZ}=\pm 1$ and with $(-,\pm)$ has $N_\bZ=\mp 1$ and $N_{M\bZ}=\pm 1$ when $m=M-k^2$ in \Cref{Hgamma}. Therefore, the numbers of the intact gapless modes in the expression of $N_\bZ$ and $N_{M\bZ}$ are written as
\be
N_{+,\pm}=\frac{N_\bZ+N_{M\bZ}}{2},\ N_{-.\pm}=\frac{N_\bZ-N_{M\bZ}}{2}.
\ee
The total number of the intact gapless edge modes in the system is the $\bZ^1$ number $N_{\bZ^1}=|N_{+,\pm}|+|N_{-,\pm}|$ as in shown \Cref{NbZ1}.

\subsubsection{Topological number of $\bZ_2$}

In \Cref{gamma1S} the $\bZ_2$ number from a system with reflection symmetry can be computed in the same way of the $\bZ_2$ number without reflection symmetry.
The reason is that the reflection symmetry does not
give rise to any new topological phase.
From the previous discussion, only for the two-copy minimal Dirac Hamiltonian ${\mH_2^d}'$ in the form of \Cref{dsigmaz} the bulk gap closing cannot be prevented by any symmetry preserving {\it homogeneous} term.
However, we will show later that this Hamiltonian can be kept gapped by
some {\it inhomogeneous} SPEMTs.

This $\bZ_2$ class can be separated into
two slightly different cases represented by the original $\bZ_2$
invariants in \cref{originaltable}:
the second descendant $\bZ_2$ of $\bZ$ in odd dimensions
and the first descendant $\bZ_2$ of $\bZ$ in even dimensions.
In the following,
we consider both of the cases together,
in the twice-as-big minimal Dirac Hamiltonian there exists a SPEMT preserving reflection symmetry ($R=i\gamma_1S$). Therefore, this reflection symmetry still keeps the original $\bZ_2$ invairants.

For the second descendant, without enlarging the dimension of the minimal Dirac Hamiltonian $d+2$ kinetic gamma matrices $\gamma_1,\gamma_2,\ldots,\gamma_{d+2}$ and one mass matrix $\gamma_0$\cite{unpublish3rd} can be used for the SPEMT construction.

For the first descendent in even
dimensions, we have the same gamma matrices, except for missing $\gamma_{d+2}$.
Therefore, from those gammas matrices some symmetry preserving terms
can be
constructed and
added into the 
enlarged minimal Dirac Hamiltonian
\begin{align}
  {\mH^d_{2}}'&=k_1\gamma_1\otimes \sigma_z+\delta\Delta+\sum_{n_i\neq 1} k_{n_i}\gamma_{n_i}\otimes \sigma_z
\nonumber \\
&\quad
+\sum_{{\rm remain}} k_{n_j}\gamma_{n_j}\otimes \bI, \label{secbz2mass}
\end{align}
where $\Delta= (i)\gamma_{d+1}\prod^{{\rm even}}_{n_i\neq 1}\gamma_{n_i}\otimes \sigma_u$
when
the first summation is over an even number of terms
or $\Delta=(i)\prod^{{\rm odd}}_{n_i\neq 1}\gamma_{n_i}\otimes \sigma_u$
when it is
over an odd number of terms.
The presence or absence of $i$ keeps $\Delta$ being hermitian and choosing $u=x, y$ lets $\Delta$ preserve TRS and PHS. We can check that $\Delta$ preserves all of the system's symmetries, including reflection symmetry $R=i\gamma_1S\otimes \bI$. The situation is similar with \Cref{deltashift}: the presence of $\Delta$ shifts the bulk gap closing points at $m=0,\ k_1=\pm \delta,\ k_{\perp}=0$. However, this system of the twice-as-big minimal Dirac Hamiltonian still can be gapped out by a special SPEMT
\begin{align}
  \hat{\frak{N}}&=
  \sum_{-\eta\leq k_1 < \eta}(ic_{k_1+\eta +\delta}\frak{N} c_{k_1-\eta+\delta}+h.c.)
  \nonumber \\
  &\quad
  +\sum_{-\eta< k_1 \leq \eta}(ic_{k_1+\eta -\delta}\frak{N} c_{k_1-\eta-\delta}+h.c.),
\end{align}
where $\mN=(i)\gamma_1\prod_{n_i\neq 1}^{{\rm even}}\gamma_{n_i}\otimes \sigma_u$
when
the first summation in \Cref{secbz2mass} is over an even number
of terms,
or
$\mN=(i)\gamma_{d+1}\gamma_1\prod_{n_i\neq 1}^{{\rm odd}}\gamma_{n_i}\otimes \sigma_u$
when the summation is over an odd number of terms.
Therefore,
$\mN$ anticommutes with all of the terms in \Cref{secbz2mass}.
Again the presence or absence of $i$
guarantees
hermiticity of $\mN$.
By
properly
choosing $u=x, y$, $\hat{\mN}$ preserves TSR and PHS by \Cref{Tgamma,Cgamma}. By the definition of the reflection symmetry operator $\hat{R}=\sum_{k_1}c^\dagger_{k_1}(i\gamma_1S\otimes \bI)c_{-k_1}$, $\hat{\mN}$ preserves the reflection symmetry. Following the similar discussion in sec.\ \ref{bz2sec}, the low energy spectrum is shown in \Cref{gapDW}. When $c>\eta$, the original bulk gap closing is blocked. No quantum phase transitions, no topological non-trivial phases. The system of the twice-as-big minimal Dirac Hamiltonian is trivial. Hence, this case is classified as $\bZ_2$ from the original topological invariant without the reflection symmetry.

\subsection{The rest of the classification}
\begin{table}
\begin{center}
\begin{tabular}{ |c | | c | c | c | c|| c|c|c|c|c|c|}
\hline			
 Class &   $TR$ &  $PH$& $Ch$  &  $R$  & d=2 & d=3 & d=4 & d=5 & d=6 & d=7  \\
\hline
\hline	
  BDI &   +1  & +1 & 1   		& $R_{-+}$ &   $0$ & $2M\bZ$ & 0 & $2\bZ$ &   $0$ & $2M\bZ$ \\
\hline
 DIII &  -1 & +1 & 1 	  		&  $R_{+-}$ &   $0$ & $2\bZ$ & 0 & $2M\bZ$ &   $0$ & $2\bZ$ \\
\hline
 CII &   -1 & -1 & 1 	  		&  $R_{-+}$ &   0 & $2M\bZ$ & $0$ & $2\bZ$ &   $0$ & $2M\bZ$  \\
\hline
CI &  +1  & -1 & 1 					 &  $R_{+-}$ &   $0$ & $2\bZ$ & 0 & $2M\bZ$ &   $0$ & $2\bZ$ \\
\hline
\end{tabular}
\caption{
\label{gammaleft}
The classification table for the rest of the commutation and anticommutation relations.
Non-trivial topology only shows up in odd spatial dimensions. }
\end{center}
\end{table}

	The classification of the rests of the commutation and anticommutation relations has to be discussed case by case. The result of this classification for the four symmetry classes is shown in \Cref{gammaleft}.

\subsubsection{Even spatial dimensions}	
We consider systems in even dimensions. In the following we show that the topological invariant in such a system is zero. For such a $d$-dimensional system, in the Dirac Hamiltonian if only $\gamma_0,\ \gamma_1, \ldots,\ \gamma_d,$ and $S$ are ingredients to construct reflection symmetry operator $R$, it is not possible that $R$ satisfies the commutation and anticommutation relations in \Cref{gammaleft}. To have $R$ from the gamma matrices, we need to introduce an extra mass term $\tilde{\gamma}$ or an extra kinetic term $\gamma_{d+1}$ preserving the non-spatial symmetries. At the same time, the Dirac Hamiltonian might be enlarged to have a new gamma matrix.
	
	With extra mass term, $R$ can be constructed and satisfy the required commutation and anticommutation relations:
\begin{align}
R_{+-}&=i\tilde{\gamma}\prod_{i=0,i\neq 1}^d\gamma_i,& {\rm as}\ &d=4n+2,  \\
R_{-+}&=\tilde{\gamma}\prod_{i=0,i\neq 1}^d\gamma_i,&\ {\rm as}\ &d=4n.
\end{align}
It is easy to check that $\tilde{\gamma}$ preserves the reflection symmetry. Hence, this gamma matrix plays a role in a SPEMT. This shows that as $R_T=\mp 1$, $R_C=\pm 1$, $d=4n(+2)$, those four symmetry classes possess the `$0$' topological invariant, which explains several $0$'s in even dimensions in \Cref{gamma1S,gammaleft}.

To have the opposite commutation and anticommutation relations in the previous case, we use $\gamma_{d+1}$ to construct
$R$
\begin{align}
R_{-+}&=i\prod_{i=0,i\neq 1}^{d+1}\gamma_i,&  {\rm as}\ d&=4n+2  \\
R_{+-}&=\prod_{i=0,i\neq 1}^{d+1}\gamma_i,&  {\rm as}\ d&=4n
\end{align}
We note that in $4n+2$ dimensions for class BDI and CII and
in $4n$ dimensions
for class DIII and CI without enlarging the Dirac Hamiltonian, an extra mass term
\be
\tilde{\gamma}_1
=\begin{cases}
 \displaystyle
S\prod_{i=0}^d\gamma_i & {\rm as}\ d=4n+2, \\
\displaystyle
iS\prod_{i=0}^d\gamma_i & {\rm as}\ d=4n,
\end{cases}
\ee
can be found.
It is to easy to check that $\tilde{\gamma}_1$ preserves system's non-spatial symmetries by \Cref{TCS}.
Furthermore, $\tilde{\gamma}_1$ anticommutes with $\gamma_{i\neq d+1}$ but commutes with $\gamma_{d+1}$ so that the reflection symmetry is preserved.
This SPEMT $\tilde{\gamma}_1$ implies this system is classified by a `$0$' original topological invariant.

\subsubsection{Odd spatial dimensions}

In odd spatial dimensions without $\tilde{\gamma}$ and $\gamma_{d+1}$, we can use only the original gamma matrices to construct $R$
satisfying the aforementioned commutation and anticommutation relations with
$T$ and $C$
\begin{align}
R_{-+}&=i\prod_{i=0,i\neq 1}^d\gamma_i,& {\rm as}\ d&=4n-1,  \label{SZleft1} \\
R_{+-}&=\prod_{i=0,i\neq 1}^d\gamma_i,& {\rm as}\ d&=4n+1.	\label{SZleft2}
\end{align}
The corresponding symmetry classes are class BDI and CII as $d=4n-1$ and class DIII and CI as $d=4n+1$. The original classification of the AZ symmetry class shows that those classes have `$0$' topological invariants so that the non-spatial SPEMT $\tilde{\gamma}$ can be present in the minimal Dirac Hamiltonian without enlarging the dimension of the matrix. However, the anticommutation relation $\{\tilde{\gamma},\gamma_i\}=0$ implies that $\tilde{\gamma}$ breaks the reflection symmetry. Such a system can be in a non-trivial phase. To find the topological invariant, we have to investigate the presence of the SPEMT in the enlarged Dirac Hamiltonian in \Cref{doublesize}.

	All of the possible ways in \Cref{doublesize} to doubling the size of the Dirac Hamiltonian can be separated into two expressions in \Cref{dbI,dsigmaz}. To construct a SPEMT, all of the ingredients are the gamma matrices, a mass term $\tilde{\gamma}$, and the Pauli matrices. We find that the first summation over only {\it odd} number of the terms in \Cref{dbI,dsigmaz} the SPEMT $\mM$ can be found. That is, $\mM=(i)\prod_{n_i\neq 1}^{{\rm odd}}\gamma_{n_i}\otimes\sigma_u$ in \Cref{dbI} and $\mM=(i)\tilde{\gamma}\gamma_1\prod_{n_i\neq 1}^{{\rm odd}}\gamma_{n_i}\otimes\sigma_u$ in \Cref{dsigmaz}, where the Hermiticity is adjusted by the presence or absence of $i$ and $u=x$ or $y$ to satisfy the non-spatial symmetries. The commutation and anticommutation $[\mM,\gamma_{n_i}\otimes\bI]=0$ and $\{\mM,\gamma_{n_j}\otimes \bI\}=0$ implies that $\mM$ commutes with the enlarged reflection symmetry operator $R'=(i)\prod_{i=0,i\neq 1}^d\gamma_i\otimes \bI$. Therefore, the reflection symmetry is preserved.  On the one hand, $\sum_{n_i\neq 1}k_{n_i}\gamma_{n_i}\otimes\sigma_z$ over an {\it odd} number of terms gives `$0$' topological invariant. On the other hand, an {\it even} number provides a non-trivial topological phase. In short, this shows that $\bZ$ topological property when $k_1$ vanishes. Such a system has the $2M\bZ(M\bZ)$\footnote{In the same dimensions the sizes of the minimal Dirac Hamiltonians in this case are always twice-as-big ones having $M\bZ$ invariants for $R_{\pm,\pm}$\cite{unpublish3rd}. For example, the size of the minimal Hamiltonian in class D$+R_{++}$ is 4 and the size of the minimal Hamiltonian is class BDI$+R_{-+}$ and CII$+R_{-+}$ is 8. The situation is similar with that in the AZ classification the sizes of the minimal Dirac Hamiltonians having $2\bZ$ invariants are twice-as-big one having $\bZ$ invariants. Hence, in this reflection symmetry case the systems possess $2M\bZ$ topological invariants.} topological invariant.

	The cases that we have not discussed are class DIII and CI in $4n-1$ dimensions and class BDI and CII in $4n+1$ dimensions. In the original classification without reflection symmetry those symmetry classes possess $\bZ$ and $2\bZ$ topological invariants. Let us first consider the $\bZ$ case. To build the reflection symmetry operator $R$ satisfying the sign changing of $R_T$ and $R_C$ in \Cref{SZleft1,SZleft2}, only using $\gamma_0,\ \gamma_1, \ldots,\ \gamma_d,$ and $S$ is not possible. Therefore, because $\sigma_y$ plays a role in switching signs of $R_T$ and $R_C$, we construct that the reflection symmetry operator is the direct product of $\sigma_y$ and the reflection symmetry operator in the form of \Cref{SZleft1,SZleft2}
\begin{align}
R_{+-}&=i\prod_{i=0,i\neq 1}^d\gamma_i\otimes\sigma_y,&  {\rm as}\ d=&4n-1,  \\
R_{-+}&=\prod_{i=0,i\neq 1}^d\gamma_i\otimes\sigma_y,&  {\rm as}\ d=&4n+1.
\end{align}
At the same time, the Dirac Hamiltonian must be enlarged in this unique way
\be
\mH^{\bZ}_2=m\gamma_0\otimes\bI+\sum_{i\neq 0}^d k_i \gamma_i\otimes \bI \label{doubleZ}
\ee
to preserve all of system's symmetries. From the properties of the $\bZ$ original topological invariant, a SPEMT is absent from this Hamiltonian. Therefore, to distinguish this non-zero topological invariant, $\mH^{\bZ}_2$ needs to be enlarged in the two possible forms
\begin{align}
\mH^{\bZ}_4=&k_1\gamma_1\otimes\bI +\sum_{n_i}k_{n_i}\gamma_{n_i}\otimes\bI\otimes \sigma_z+\sum_{n_j}k_{n_j}\gamma_{n_j}\otimes\bI_{4\times 4}, \label{bigbI} \\
\mH'^{\bZ}_4=&k_1\gamma_1\otimes\sigma_z +\sum_{n_i}k_{n_i}\gamma_{n_i}\otimes\bI\otimes \sigma_z+\sum_{n_j}k_{n_j}\gamma_{n_j}\otimes\bI_{4\times 4}. \label{bigZ}
\end{align}
Again the original $\bZ$ protects the non-trivial topological phases when an even number of $\sigma_z$ in the Hamiltonians. For the case of an odd number, although an extra mass term, which preserves the non-spatial symmetries, is present, we have to confirm that the reflection symmetry is also preserved so that the system is in the trivial phase. The expression of the mass term is $\mM=(i)\prod_{n_i\neq 0}^{{\rm odd}}\gamma_{n_i}\otimes \sigma_x\otimes\sigma_u$ for \Cref{bigbI} and $\mM=(i)\gamma_0\prod_{n_i\neq 0}^{{\rm odd}}\gamma_{n_i}\otimes \bI\otimes\sigma_u$ for \Cref{bigZ}, where the presence or absence of $i$ keeps the Hermiticity of $\mM$ and choosing $u=x$ or $y$ makes $\mM$ preserve the non-spatial symmetries. It is easy to check that $\mM$ commutes with $R=(i)\prod_{i=0,i\neq 1}^d\gamma_i\otimes\sigma_y\otimes\bI$. Hence, the behavior of even and odd numbers of $\sigma_z$ shows that even with reflection symmetry such a system still possesses $\bZ$ properties. However, the reflection symmetry requires the dimension of the Dirac Hamiltonian to be doubled in \Cref{doubleZ}. The classification of the topological invariant changes from $\bZ$ to $2\bZ$.

For the $2\bZ$ case, we know that in the proper basis the minimal Dirac Hamiltonian of $2\bZ$ is the two copies of the minimal Dirac Hamiltonian of $\bZ$:
\be
\mH^{2\bZ}=m\gamma_0^\bZ\otimes\bI+\sum_{i\neq 0}^d k_i \gamma_i^\bZ\otimes \bI.
\ee
The non-spatial symmetry operators can be expressed by the symmetry operator of $\bZ$
\be
T^{2\bZ}=T^{\bZ}\otimes \sigma_y,\ C^{2\bZ}=C^{\bZ}\otimes \sigma_y.
\ee
Following the similar discussion of the $\bZ$ case, a system of the $2\bZ$ original topological invariant with the reflection symmetry is still classified as $2\bZ$.

\section{Examples}
\label{examples}

Let us now discuss several examples
of topological phases protected by reflection symmetry:
more specifically, we will consider:
\begin{description}
\item[class AIII$+R_+$ and BDI$+R_{++}$ in $d=2$]
All of the non-spatial symmetry operators commutes with $R$. Therefore,  \Cref{commutationtable} shows that both of the symmetry classes possess $M\bZ$ topological invariants.

\item[class DIII$+R_{--}$ in $d=2$]
In this symmetry class,
$R$ anticommutes
with $T$ and $C$. This situation hence falls into
\Cref{anticommutationtable}. As described below,
gapped phases in this case
are characterized an $M\bZ$ topological invariant.

\item[class D$+R_{+}$ in $d=2$]
With reflection symmetry,
gapped phases in this symmetry class
are characterized as a $\mathbb{Z}_2$ topological invariant
in \Cref{commutationtable}.

\item[class CII$+R_{--}$ in $d=2$]
For class CII no non-zero topological invariants are in the original classification or for \emph{mirror} symmetry class A  in the reflection symmetric plane. The gapped phases are always trivial.

\item[class DIII$+R_{-+}$ in $d=3$] $R$ anticommutes with $T$ but commutes with $C$. Hence, \Cref{gamma1S} shows there is a $\bZ^1$ in this case.

\end{description}

\subsection{Class AIII+$R_+$, BDI+$R_{++}$, and DIII+$R_{--}$ in $d=2$}
\label{AIII+R and BDI+R in $d=2$}

Symmetry class AIII has two physical interpretations,
one in terms of charged (complex) fermions
with conserved fermionic number,
and
the other in terms of BdG Hamiltonians.
As an electron system, a way to obtain such a system is to consider
lattice fermion systems with
bipartite hopping only.
In this context, chiral symmetry of class AIII is {\it sublattice symmetry}.
Alternatively,
symmetry class AIII can be realized
as a time-reversal symmetric BdG Hamiltonian
with conserved $S_z$ spin rotation.
While it is perhaps fair to say that the BdG interpretation is more
experimentally realizable,
since achieving an exact sublattice symmetry is challenging,
in this section we focus on electronic realizations of symmetry class AIII.

Similarly to class AIII,
symmetry class BDI has two physical interpretations,
one in terms of charged (complex) fermions
with conserved fermionic number,
and
the other in real (Majorana) fermions.
Below, we will first discuss realization in terms of
complex fermions; we will later discuss
a realization in terms of Majorana fermions.

Recall that in $d=2$,
there is no topological insulator in AIII and BDI
if we do not impose reflection symmetry.
Class DIII is of $\bZ_2$ type.
With reflection, there are topological insulators
in these three classes
characterized by $M\bZ$ topological invariants.

\paragraph{bulk Hamiltonian}

Let us start by considering
the following tight-binding Hamiltonian:
\begin{align}
H
&=
\sum_{r}
\psi^{\dag}(r)
\left(
\begin{array}{cc}
t & {i}\Delta\\
{i}\Delta  & -t
\end{array}
\right)
\psi(r+\hat{x})
+\mathrm{h.c.}
\nonumber \\
&
\qquad
+
\psi^{\dag}(r)
\left(
\begin{array}{cc}
t & \Delta\\
-\Delta  & -t
\end{array}
\right)
\psi(r+\hat{y})
+\mathrm{h.c.}
\nonumber \\
&
\qquad
+
\psi^{\dag}(r)
\left(
\begin{array}{cc}
\mu & 0\\
0  & -\mu
\end{array}
\right)
\psi(r),
\label{lattice p-wave}
\end{align}
where
the two-component fermion annihilation
operator at site $r$,
$\psi(r)$,
is given in terms of
the electron annihilation operators
with spin up and down,
$c_{r,1/2}^{\ }$,
as
$\psi^{T}({r})=
(c_{r,1}, c_{r,2})$,
and we take
$t=\Delta=1$ and $\mu=m+2$.
The chiral
$p$-wave superconductor
has been discussed in the context of
superconductivity in strontium ruthenate\cite{Kallin:2009fk}
and paired states in the fractional quantum Hall effect\cite{PhysRevB.61.10267}.
There are four phases separated by three quantum critical points
at $\mu=0,\pm 4$, which are labeled by
the Chern number as
$\mathrm{Ch}=0$ $( |\mu| > 4)$,
$\mathrm{Ch}=-1$ $(-4 < \mu < 0)$, and
$\mathrm{Ch}=+1$ $( 0 < \mu < +4)$.
The non-zero Chern number implies the IQHE in the spin transport.
\cite{Senthil1998}
In momentum space,
\begin{align}
H
&=
\sum_{k \in \mathrm{BZ}}
\psi^{\dag}(k)
\left[
\vec{n}(k) \cdot\vec{\sigma}
\right]
\psi(k),
\nonumber \\
\vec{n}(k)
&=
\left(
\begin{array}{c}
-2 \Delta \sin k_x \\
-2 \Delta \sin k_y \\
2t (\cos k_x + \cos k_y) + \mu
\end{array}
\right).
\end{align}

A lattice model topological insulator
in symmetry class
AIII $+R_+$,
BDI $+R_{++}$,
and
DIII $+R_{--}$
can be constructed by
taking the two copies of the above two-band Chern insulator
with opposite chiralities.
Consider the Hamiltonian in momentum space,
\begin{align}
H
&=
\sum_{k \in \mathrm{BZ}}
\sum_{s=\uparrow,\downarrow}
\psi^{\dag}_{s}(k)
\left[
\vec{n}_{s}(k) \cdot\vec{\sigma}
\right]
\psi_{s}(k),
\label{chiral+R}
\end{align}
where
$s=\uparrow,\downarrow$
represent ``pseudo spin'' degrees of freedom,
and
$\vec{n}_{s}(k)$ is given, in terms of $\vec{n}$ as
\begin{align}
\vec{n}_{\uparrow}(k)
&=
\vec{n}(k),
\quad
\vec{n}_{\downarrow}(k)
=
\vec{n}_{\uparrow}(\tilde{k})
=
\vec{n}(\tilde{k})
,
\end{align}
where $\tilde{k} = (-k_1, k_2, \ldots) = (-k_1, k_{\perp})$.
I.e.,
\begin{align}
\mathcal{H}
(k)
=
n_x(k) \tau_z \sigma_x
+
n_y(k) \tau_0 \sigma_y
+
n_z(k) \tau_0 \sigma_z.
\end{align}
The model is chiral symmetric:
\begin{align}
S^{-1} \mathcal{H}(k) S
=
-
\mathcal{H}(k),
\quad
S = \tau_x \sigma_x.
\end{align}
The Hamiltonian is invariant under the following two TRS:
\begin{align}
T^{-1} \mathcal{H}(-k) T
=
\mathcal{H}(k),
\quad
T = \tau_x \sigma_0\Theta,
\quad
T^2 =+ 1
\nonumber \\
T^{-1} \mathcal{H}(-k) T
=
\mathcal{H}(k),
\quad
T = \tau_y \sigma_0\Theta,
\quad
T^2 = -1.
\end{align}
Also, the corresponding particle-hole symmetries with $C=ST$ are
\begin{align}
C^{-1} \mathcal{H}(-k) C
=&
-\mathcal{H}(k),
\quad &
C =& \tau_0 \sigma_x\Theta,
\quad &
C^2 =&+ 1
\nonumber \\
C^{-1} \mathcal{H}(-k) C =&
-\mathcal{H}(k),
\quad &
C =& i\tau_z \sigma_x\Theta,
\quad &
C^2 =&+ 1.
\end{align}
Imposing the former form of TRS,
the system falls into symmetry class BDI,
whereas
with the latter form of TRS,
the system falls into symmetry class DIII.

We now impose the following reflection symmetry:
\begin{align}
R^{-1} \mathcal{H}(\tilde{k}) R
=
\mathcal{H}(k),
\quad
R = \tau_x.
\end{align}
Chiral and reflection symmetries commute
with each other in the sense that
\begin{align}
[S, R]=0.
\label{r and chiral commute}
\end{align}
For class BDI, $R$ commutes with $T$ and $C$ ($R_{++}$). For class DIII, $R$ anticommutes with $T$ and $C$ ($R_{--}$). From \Cref{commutationtable,anticommutationtable}, both of the cases possess $M\bZ$ topological invariants.
Similarly, class AIII$+R_{+}$ is also classified by a $M\bZ$ topological invariant.

\paragraph{$M\bZ$ bulk topological invariant}\label{windingcompute}

At the reflection symmetric plane,
i.e., only $k_x=0$ in the continuum model,
$\mathcal{H}(k)$
commutes with $R$, and hence,
it can be block diagonalized.
Furthermore,
since reflection $R$ commutes with chiral symmetry
in \Cref{r and chiral commute},
each block has an off-diagonal structure in a proper basis,
\begin{align}
\mathcal{H}(0,k_y)
&=
\left(
\begin{array}{cc}
\mathcal{H}^+(0,k_y) & 0\\
0 & \mathcal{H}^-(0,k_y)
\end{array}
\right),
\nonumber \\
\mathcal{H}^{\pm}(0, k_y)
&=
\left(
\begin{array}{cc}
0 & D_{\pm}(k_y) \\
D^{\dag}_{\pm}(k_y) & 0
\end{array}
\right),
\end{align}
where $\pm$ indicates the eigenspace of $R=\pm 1$.
For each block,
the 1D winding number
(the topological invariant of symmetry class AIII
in $d=1$)
is well-defined
\cite{Schnyder:2008gf}; by the definition of the $M\bZ$ number in \Cref{numSZ}, we only need to focus on the winding number in one of the eigenspaces of $R$, say $+1$, due to the absence of the weak index. The winding number is defined from $q(k)$ in the $Q$-matrix in \Cref{Qmatrix}. To construct the $Q$-matrix from the ($k_y$-dependent) occupied wavefunctions, we have to solve the eigenvalue problem
\be
\mH^+\Phi^a_\pm=\pm\varepsilon^a\Phi^a_\pm,
\ee
where $a$ runs over occupied bands, $a=1,2$ in our case and
\be
\Phi^a_\pm=\frac{1}{\sqrt{2}}
\bma u^a \\ \pm v^a \ema^T.
\ee
Here we assume $u^a$ and $v^a$ are normalized so $\Phi^a_\pm$ is also normalized. Because $(\mH^+)^2\Phi^a_\pm=(\varepsilon^a)^2\Phi^a_\pm$, $u^a$ and $v^a$ are the eigenfunctions of $D^{\ }_{R} D^{\dag}_{R}$
and
$D^{\dag}_{R} D^{\ }_{R}$ respectively and share the same positive eigenvalue $(\varepsilon^a)^2$:
\begin{align}
D  D^{\dag}
u^a =
(\varepsilon^a)^2 u^a,
\quad
D^{\dag} D^{\ }
v^a =
(\varepsilon^a)^2
v^a.
\end{align}
Therefore, we can compute the projector of the occupied bands which have negative energies
\begin{align}
P(k_y)=&\frac{1}{2}\sum_a
\bma
u_a \\ -v_a
\ema
\bma
u_a^\dagger & -v^\dagger_a
\ema \nonumber \\
=&
\frac{1}{2}\bma
\bI & 0 \\
0  & \bI \\
\ema
-
\frac{1}{2}\sum_a
\bma
0 & u_av^\dagger_a \\
v_au^\dagger_a & 0 \\
\ema.
\end{align}
Therefore, due to $Q(k_y)=\bI-2P(k_y)$ and the definition of $q(k_y)$ in \Cref{Qmatrix}, we have the expression of
\begin{align}
q(k_y)
=
\sum_a
| u^a(k_y)\rangle \langle v^a(k_y)|,
\end{align}
the topological invariant for $\mH^+$ at $k_x=0$ is defined by
\begin{align}
\nu_{0}
=
\frac{i}{2\pi }\int dk_y\,
\mathrm{tr}\,
\big[
q^{\dag}(k_y) \partial_{k_y} q(k_y)
\big].
\end{align}
The winding number $\nu_\pi$ at the other symmetric point $k_x=\pi$ can be computed in the similar way so we have the $M\bZ$ number $N_{M\bZ}$ by \Cref{numSZ}.

Following the general discussion,
we now calculate the topological
invariant of the model in \Cref{chiral+R}.
We first look for a basis where
(i) $R$ is diagonal,
(ii) $\mathcal{H}(k)$ is block-diagonal for each
reflection eigenvalue $R=\pm 1$,
and
(iii)
chiral symmetry operation looks identical for
both sectors of $R$.
This can be achieved by the unitary transformation
$U=U_1U_2$ where
\begin{align}
U_1 &=
\frac{1}{\sqrt{2}}
(
\tau_0 \sigma_0
+
\tau_z \sigma_y
),
\quad
U_2
=\frac{1}{\sqrt{2}} (\tau_0  + {i}\tau_y)
.
\end{align}
Under the unitary transformation,
\begin{align}
U R U^{\dag}
&=
\tau_z \sigma_0,
\quad
U S U^{\dag}
=
\tau_0 \sigma_z,
\nonumber \\
U \mathcal{H}(K) U^{\dag}
&
=
\left(
\begin{array}{cc}
 \mathcal{H}^+(K) & 0 \\
0 & \mathcal{H}^- (K) \\
\end{array}
\right),
\end{align}
where $K=0,\ \pi$, and
\begin{align}
\mathcal{H}^{\pm}_{K}(k_y)
&=
\left(
\begin{array}{cc}
0 & -i n_y \mp n_z \\
i n_y \mp n_z & 0
\end{array}
\right).
\end{align}
The topological invariant $\nu^{\pm}_K$
for the blocks
$\mathcal{H}^{\pm}_K$
satisfy
$\nu^{+}_K
=
-
\nu^{-}_K
$ due to the absence of the weak index. In this specific model, $\nu_0^+=1$ and $\nu_\pi^+=0$ as $-4<\mu<0$ and $\nu_0^+=0$ and $\nu_\pi^+=1$ as $0<\mu<4$. Therefore, by \Cref{numSZ} $N_{M\bZ}=1$ and $-1$ as $-4<\mu<0$ and $0<\mu<4$ respectively.

\paragraph{edge theory}

Let us now introduce a boundary to the system.
In the continuum model, one way to do this is to
make the mass $y$-dependent
$m\to m(y)$.
The edge Hamiltonian, in a suitable basis, is
\begin{align}
\mathcal{H}(k_x) = k_x \sigma_3,
\quad
\{ \mathcal{H}(k_x), \sigma_1\} =0.
\end{align}
Or, in the second quantized language,
the edge mode is described by
\begin{align}
H
 &=
 \int dx
 \left[
 \psi^{\dag}_L {i}\partial_x \psi^{\ }_L
-
 \psi^{\dag}_R {i} \partial_x \psi^{\ }_R
 \right].  \label{LRchiral}
\end{align}
Since there are left and right movers,
one could give a mass to gap them out.
We can write down such two masses,
\begin{align}
m(\psi_L^{\dag} \psi^{\ }_R + \psi_L^{\dag} \psi^{\ }_R),
\quad
{i}m_5 (\psi_L^{\dag} \psi^{\ }_R - \psi_L^{\dag} \psi^{\ }_R),
\end{align}
if we have not imposed any discrete symmetry.
Let us now impose the two discrete symmetries.
With chiral symmetry, the first mass will be eliminated:
the first mass term can be written
as
$m\Psi^{\dag} \sigma_x \Psi$,
where
$\Psi:=
(
\psi_L,
\psi_R
)^T.
$
Since $\{m \sigma_x, \sigma_x\}\neq 0$,
this mass term is not compatible with chiral symmetry.
On the other hand, the second mass is
$m_5 \Psi^{\dag} \sigma_y \Psi$.
Since $\{m_5 \sigma_y, \sigma_x\}= 0$,
this mass term is chiral symmetric, and allowed to exist.
With reflection symmetry,
we should be able to prohibit the second mass term.
By definition, reflection should exchange $\psi_L$ and $\psi_R$,
and observing reflection should commute with chiral symmetry,
$R$ is given by
\begin{align}
\mathcal{R}\psi_L(x)\mathcal{R}^{-1}
= \psi_R(-x),
\quad
\mathcal{R}\psi_R(x)\mathcal{R}^{-1} =\psi_L(-x).
\end{align}
Observe that
reflection and chiral symmetry commute,
\begin{align}
[R,\sigma_x] = 0.
\end{align}
The first mass as well as
the kinetic term
is invariant under $\mathcal{R}$.
However, the second mass
is not invariant under $\mathcal{R}$.
Therefore, with both chiral and
reflection symmetries,
the edge state is stable.

We have treated the case with
unit topological invariant.
By increasing the number of edge channels with the same or different signs in \Cref{LRchiral}, any mass term is still prohibited.
We can consider cases with $M\bZ$ topological invariants.

\subsection{Class D+$R_{+}$ in $d=2$}

\label{class D+R (and BDI+R) in $d=2$}

For symmetry class D
(i.e.,\ generic BdG systems
without any symmetry)
in $d=1$,
the system is classified as a $\mathbb{Z}_2$ topological superconductor. Therefore, a $\mathbb{Z}_2$ topological superconductor can be realized in a symmetry class D system with reflection symmetry $R_{++}$ due to the upward shifting in \Cref{commutationtable}.
As follows, we will describe this case in more details with an example.

Observe that the pairing terms in $x$-direction
for the spin up and spin down sector differ
by sign.
The model is invariant under reflection defined by
\begin{align}
\mathcal{R} c_{s r} \mathcal{R}^{-1}
=
c_{-s \tilde{r}}
\quad
(s=\uparrow, \downarrow).
\end{align}
Furthermore, for spin-one-half system, $R^2=-1$. Without affecting the equation above, we performs a phase shift so that $R=\sigma_x$ is hermitian. In particular,
note that
the pairing terms in $x$-direction
transform under reflection $\mathcal{R}$ as
$\mathcal{R}$:
$
\sum_r
\Delta (
c_{\uparrow{r}}^{\dag}
c_{\uparrow{r}+\hat{{x}}}^{\dag}
-
c_{\uparrow{r}+\hat{{x}}}^{\dag}
c_{\uparrow{r}}^{\dag}
)
\to
-\sum_r
\Delta  (
c_{\downarrow{r}}^{\dag}
c_{\downarrow{r}+\hat{{x}}}^{\dag}
-
c_{\downarrow{r}+\hat{{x}}}^{\dag}
c_{\downarrow{r}}^{\dag}
)
$,
and hence
$\mathcal{R}$
exchanges $H^{\uparrow}_{p+{i}p}$
and $H^{\downarrow}_{-p+{i}p}$.

\paragraph{bulk spectrum}

With the periodic boundary condition,
we make use of the Fourier transforms
which transform the Hamiltonian
into
\begin{align}
&
H
=
\sum_{0\le k_x \le \pi}
\sum_{k_y}
\Psi^{\dag}_{k_x}(k_y)
\mathcal{H}_{k_x}(k_y)
\Psi^{\ }_{k_x}(k_y),
\nonumber \\
&\Psi^{\dag}_{k_x}(k_y)
: =
\big(
\begin{array}{cccc}
c^{\dag}_{\uparrow,k}, &
c^{\ }_{\downarrow,-k}, &
c^{\dag}_{\uparrow,k}, &
c^{\ }_{\downarrow,-k}
\end{array}
\big).
\end{align}
where
the kernel
$\mathcal{H}_{k_x}(k_y)$
is block diagonal in spin indices and given by
\begin{align}
\mathcal{H}_{k_x}(k_y)
&=
\left(
\begin{array}{cc}
\mathcal{H}^{\uparrow}_{k_x}(k_y) & 0 \\
0 & \mathcal{H}^{\downarrow}_{k_x}(k_y)
\end{array}
\right),
\nonumber \\
\mathcal{H}^{s}_{k_x}(k_y)
&=
\left(
\begin{array}{cc}
\xi_{k} & \Delta^s_{k} \\
\Delta^{s*}_{k} & -\xi_{k}
\end{array}
\right),
\end{align}
where $\xi_k
=
2t (\cos k_x+\cos k_y)-\mu$ and $\Delta^s_k
=
2\Delta(-  s \sin k_x + {i}\sin k_y)$. The particle-hole symmetry operator $C=\tau_x\Theta$ commutes with $R=\sigma_x$. It corresponds to class D$+R_{++}$. At the reflection symmetric points $k_x=0,\ \pi$,
following the general discussion,
we take a basis in which $R$ is diagonal:
this is achieved by a unitary transformation,
\begin{align}
R
&\to
U R U^{-1}
=
\sigma_z,
\quad
U = \frac{1}{\sqrt{2}}
\left(
\begin{array}{cc}
1 & 1 \\
1 & -1
\end{array}
\right).
\end{align}
Accordingly, the BdG Hamiltonian
at $k_x=0,\pi$,
$\mathcal{H}_{k_x=0,\pi}(k_y)$,
is transformed as
\begin{align}
\mathcal{H}_{k_x}(k_y)
&\to
\left(
\begin{array}{cc}
\mathcal{H}^+_{k_x}(k_y) & 0 \\
0 & \mathcal{H}^-_{k_x}(k_y)
\end{array}
\right),
\nonumber \\
\mathcal{H}^{R=\pm}_{k_x}(k_y)
&=
\frac{1}{2}
\left[
\mathcal{H}^{\uparrow}_{k_x}(k_y)
+
\mathcal{H}^{\downarrow}_{k_x}(k_y)
\right],
\end{align}
where note
$\mathcal{H}^{\uparrow}_{k_x=0,\pi}(k_y)
=\mathcal{H}^{\downarrow}_{k_x=0,\pi}(k_y)
$
in this example.
The Hamiltonian for $R=+1$ and $R=-1$ sectors
are identical, and given by
\begin{align}
&
\mathcal{H}^{\pm}_{k_x=0,\pi}(k_y)
=
\left(
\begin{array}{cc}
\xi_{k_y} & \Delta_{k_y} \\
\Delta_{k_y}^* & -\xi_{k_y}
\end{array}
\right),
\nonumber \\
&
\xi_{k_y}
=
2t (\pm 1 +\cos k_y)-\mu,
\quad
\Delta_{k_y}
=
2{i} \Delta\sin k_y,
\label{ref H}
\end{align}
where $+$ for $k_x=0$ and $-$ for $k_x=\pi$.

\paragraph{bulk topological invariant}

Once we decompose the Hamiltonian
into reflection symmetric/antisymmetric ($R=\pm $) sectors,
we can define a $d=1$ topological invariant
for each $R=\pm $ sector,
following Ref.\ \onlinecite{Kitaev1D}:
Let us consider class D systems in $d=1$ with translation
invariance.
In momentum space, it can be written as
\begin{align}
H = \frac{{i}}{4}
\sum_{\alpha,\beta} \sum_{k_y}
\tilde{B}_{\alpha\beta}(k_y)
\lambda_{\alpha}(k_y)
\lambda_{\beta}(-k_y),
\end{align}
where $\alpha,\beta$ are a band index,
and
we are using a Majorana fermions $\lambda_{\alpha}(k_y)$
to represent a class D system.
Then the $\mathbb{Z}_2$ topological invariant is given by
\begin{align}
(-1)^{\nu}
:=
\mathrm{sgn}\big[\mathrm{Pf}\, \tilde{B}(0) \big]
\mathrm{sgn}\big[\mathrm{Pf}\, \tilde{B}(\pi) \big]
=
\pm 1.
\end{align}

In our problem,
this Kitaev $\mathbb{Z}_2$ invariant
can be defined for $k_x=0,\ \pi$
and for each $R=\pm$ sector.
For a given sector ($R=+$, say), by performing the transformation to real Majorana fermion operators
\begin{align}
\label{maj:eq1}
\left\{
\begin{array}{l}
\lambda_{k} := c_{k_y}^{\dag}+ c_{k_y}^{\ },
\\
\lambda'_{k} := (c_{-k_k}^{\ }- c_{-k_y}^{\dag})/{{i}},
\end{array}
\right.
\quad
\Lambda_k
:=
\left(
\begin{array}{c}
\lambda_{k_y} \\
\lambda'_{k_y}
\end{array}
\right),
\end{align}
the Hamiltonian in \Cref{ref H}
is given by
\begin{align}
H^R_{k_x=0,\pi}(k_y)
&=
\frac{-{i}}{4}
\sum_{{k}}
\Lambda^T_{-k_y}
\bma
2\Delta \sin k_y & -\xi_{k_y} \\
\xi_{k_y} & 2\Delta \sin k_y \\
\ema
\Lambda_{k_y}.
\end{align}
Let us focus on the $\bZ_2$ topological invariant at $k_x=0$. Thus, the $\tilde{B}$-matrix at $k_y=0,\ \pi$ is given by
\begin{align}
\tilde{B}_{0}(0)
&=
\left(
\begin{array}{cc}
0 & 4t-\mu \\
-4t+\mu & 0
\end{array}
\right),
\tilde{B}_{0}(\pi)
=
\left(
\begin{array}{cc}
0 & -\mu \\
+\mu & 0
\end{array}
\right),
\end{align}
and hence the topological invariant is
\begin{align}
(-1)^{\nu_0}&=\mathrm{sgn}\big[\mathrm{Pf}\, \tilde{B}_0(0) \big]
\mathrm{sgn}\big[\mathrm{Pf}\, \tilde{B}_0(\pi) \big] \nonumber \\
&=\mathrm{sgn}(4t-\mu)\cdot\mathrm{sgn}(-\mu)
\end{align}
Similarly, the topological invariant at $k_x=\pi$ is
\be
(-1)^{\nu_\pi}=\mathrm{sgn}(-4t-\mu)\cdot\mathrm{sgn}(-\mu)
\ee
Therefore, the $\bZ_2$ topological invariant in \Cref{Z2def} for the entire reflection system is
\begin{align}
(-1)^{N_{\bZ_2}}
&=
 \mathrm{sgn}\,(+4t -\mu )
\cdot  \mathrm{sgn}\,(- 4t-\mu).
\end{align}
The presence of the topological nontrivial phase is in the region $-4t<\mu<4t$.

\paragraph{edge state}

A consequence of non-trivial topology in the bulk
is the presence of edge modes. Let us now consider an edge of $d=2$ class D topological
superconductor:
\begin{align}
S
=
\int dxd\tau
\left(
\psi_{\uparrow} (-i\partial_x) \psi_{\uparrow}
+
\psi_{\downarrow} (i\partial_x) \psi_{\downarrow}
\right),
\end{align}
where $\psi_{\uparrow}$ and $\psi_{\downarrow}$
are a left-moving (right-moving) Majorana fermion.
A mass term
$
{i} m\psi_{\uparrow}\psi_{\downarrow},
$
is prohibited because of a reflection symmetry:
\begin{align}
R: \psi_{\uparrow} \to \psi_{\downarrow},
\quad
\psi_{\downarrow} \to \psi_{\uparrow}.
\end{align}
In other words, with reflection the edge state is stable,
and there is a bulk topological superconductor protected
by reflection.

Such protection by reflection symmetry is of $\mathbb{Z}_2$ type
as we can readily see by considering
two copies of the edge theory:
\begin{align}
S
=
\int dxd\tau
\sum_{i=1}^2
\left(
\psi_{\uparrow i} (-i\partial_x) \psi_{\uparrow i}
+
\psi_{\downarrow i} (i\partial_x) \psi_{\downarrow i}
\right).
\end{align}
 The corresponding Hamiltonian is
\begin{align}
H
&= \int dx\, \Psi^T \mathcal{H} \Psi,
\quad
\mathcal{H}
=
-{i} \partial_x  \tau_0 \sigma_z.
\end{align}
where
$\Psi^T = (\psi_{\uparrow 1}, \psi_{\downarrow 1}, \psi_{\uparrow 2},\psi_{\downarrow 2})^T$.
Mass terms
$
{i}\psi_{\uparrow 1}\psi_{\downarrow 1}
$
and
$
{i} \psi_{\uparrow 2}\psi_{\downarrow 2}
$
are again prohibited.
However,
$-2 {i}  (\psi_{\uparrow 1} \psi_{\downarrow 2}  +\psi_{\downarrow 1}\psi_{\uparrow 2})
=
 \Psi^T \tau_y \sigma_x \Psi,
$
and
$-2 {i}  (\psi_{\uparrow 1} \psi_{\uparrow 2}  +\psi_{\downarrow 1} \psi_{\downarrow 2})
=
 \Psi^T \tau_y \sigma_0 \psi,
$
are invariant under reflection and hence allowed to be added
as a perturbation.


\subsection{Class CII+$R_{--}$ in $d=2$}\label{egfakeZ2}
\begin{figure}
\begin{center}
\subfigure[Real spectrum: size $L_x\times L_y=140\times 24$ ]{\label{realsp}\includegraphics[width=70mm]{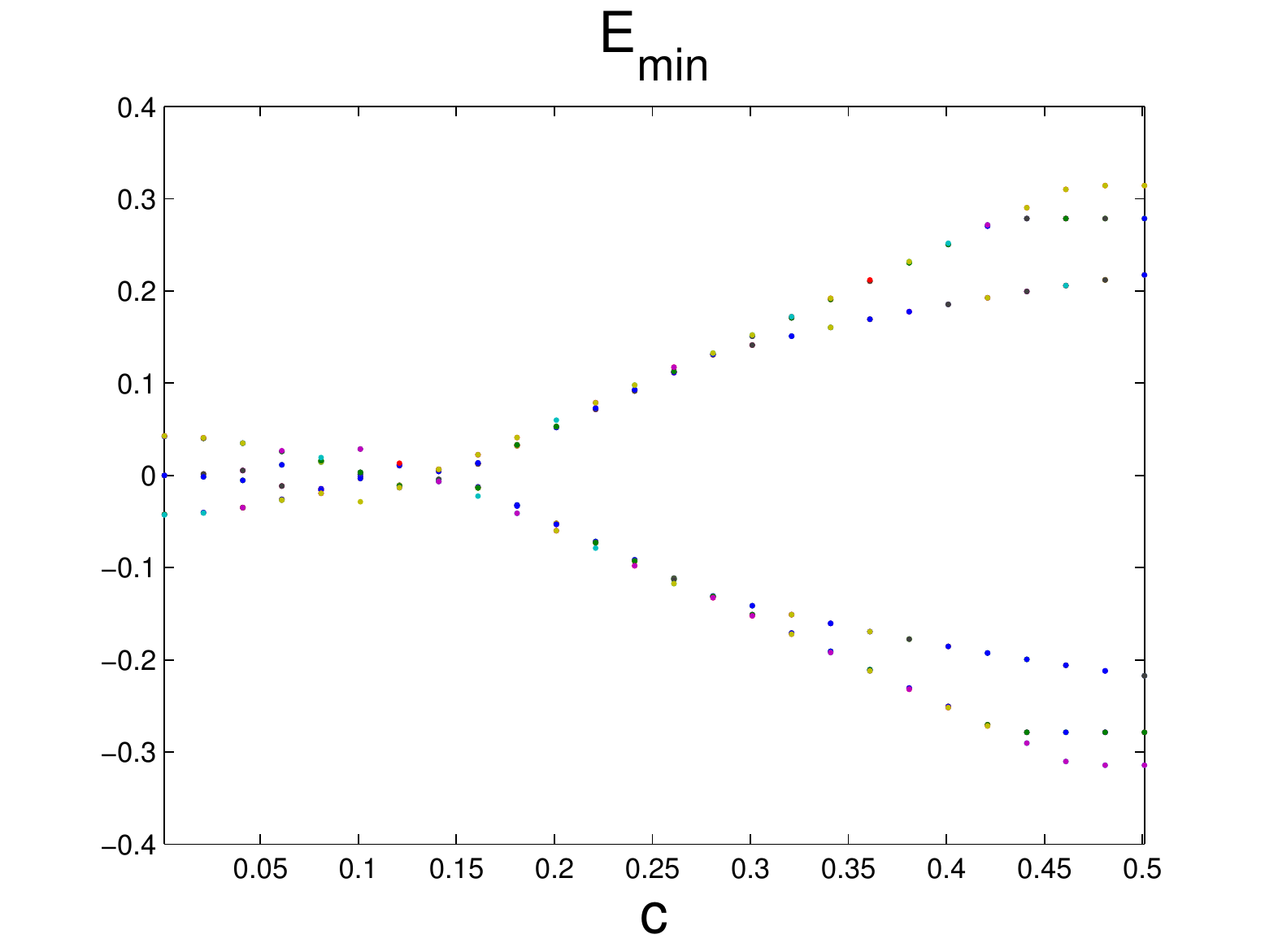}}
\subfigure[Size $L_x\times L_y=140\times 8$]{\label{Entsp}\includegraphics[width=70mm]{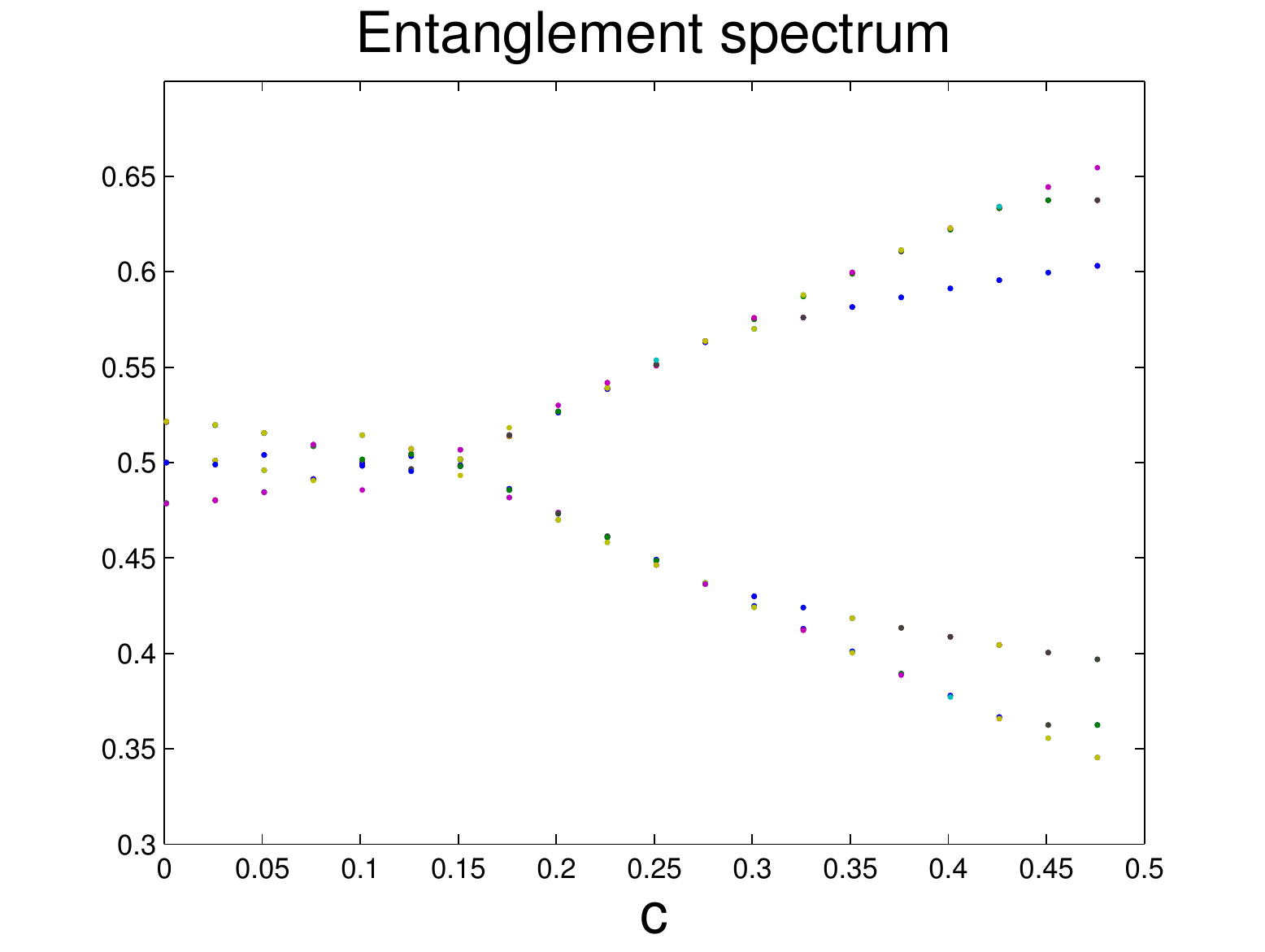}}
\end{center}
\caption{
 (a) The energy spectrum for the $y$ edge states near $E=0$ as a function of
 a translational symmetry breaking term ($c$),
 and (b)
 the entanglement spectrum around the entanglement eigenvalue 1/2
 as a function of $c$.
 We consider the open boundary condition is the $y$ discretion.
 The parameters are set to $m=-1$, $\delta=\arcsin(0.1\pi)$, and $\eta=0.05\pi=0.157$.
 As $c>0.157$, the edge states start being gapped and the entanglement mid-gap states
 begin to move out from $0.5$.
 However, as $c<0.157$,
 $E_{{\rm min}}$ is nonzero and the entanglement eigenvalue is not $0.5$ due to the finite size effect in the numerical stimulation.}
\label{gapped}
\end{figure}

In class CII a system is invariant under TS, PH with $T^2=C^2=-1$.
When we impose reflection symmetry into the system and require
$R$ to anticommute with $T$ and $C$, \Cref{anticommutationtable} shows that class CII+$R_{--}$ in 2 dimension is always in the trivial phase. 
In this section, we will provide one example to show that the gapless edge states and the entanglement mid-gap states inevitably are gapped by the translational breaking term, which preserves all of the discrete symmetries, in \Cref{NDW}. Consider the Hamiltonian in class CII+$R_{--}$
\be
\mH=M\gamma_0+\sin k_1 \gamma_1+ \sin k_2 \gamma_2,
\ee
where $M=m+\cos k_1+\cos k_2$, $\gamma_0=\tau_z$, $\gamma_1=\tau_x\six$, and $\gamma_2=\tau_x\siy\mu_x$.
The corresponding non-spatial symmetry operators are $T=\siy\Theta$, $C=\tau_x\mu_y\Theta$. Physically, $\tau_i$, $\sigma_j$, and $\mu_l$ can be treated as particle-hole, one-half spin, and pseudo-spin degrees of freedom respectively. Without enlarging the matrix dimension, we have extra three kinetic terms $\gamma_3=\tau_x\siz,\ \gamma_4=\tau_y,$ and $\gamma_5=\tau_x\siy\mu_z$.
In addition, $\sin k_i\gamma_{3,4,5}$ preserve TRS and PHS.
Therefore, the reflection symmetry operator
can be defined as $R=i\gamma_1\gamma_3=\siy$,
which anticommutes with $T$ and $C$.
For such a reflection system, as $-2<m<2$ the topological phase is non-trivial if the translational symmetry is preserved. That is, the real spectrum gapless edge states still can be observed at the $y$ direction edge and the entanglement mid-gap states are present when we make a spatial cut exactly in half at $x=0$. However, those states are unstable when the translational symmetry is broken. Consider the Hamiltonian $\mH$ with $\delta\Delta=i\delta\gamma_1\gamma_{3}\gamma_{4}=\tau_y\siy$ and a translational symmetry breaking term $c\hat{\mN}$ in \Cref{NDW}, where $\mN=i\gamma_{d+1}\gamma_{d+2}\gamma_{d+3}=\tau_y\six\mu_z$.  As shown in \Cref{realsp} the edge states are gapped when $c>\eta$. Furthermore, \Cref{Entsp} shows that the mid-gap in the entanglement spectrum is destroyed; this confirms that in the region $-2<m<2$ the system is in the trivial phase.

\subsection{Class DIII+$R_{-+}$ in $d=3$ }
By reading \Cref{gamma1S}, in three spatial dimensions class DIII$+R_{-+}$ possesses $\bZ^1$. In what follows, we demonstrate $\bZ^1$ topological properties by considering this reflection symmetric system in class DIII in three dimensions. First, without reflection symmetry we introduce a three-dimensional time reversal invariant topological insulator possessing \emph{artificial} chiral symmetry corresponding to class DIII. We choose the Hamiltonian of such a system to be identical to the Bogoliubov-de Gennes Hamiltonian of $^3$He-B phase\cite{review_TIb,Chung:2009fk}. The system of $^3$He-B is recently shown to be a topological superfluid preserving TRS and PHS  in three dimensions\cite{Wada:2008ly,Murakawa:2009ve,JPSJ.80.013602}. Now instead of PHS, we require our system to be invariant under chiral symmetry even if TRS is broken. Therefore, we write down the Hamiltonian of the 3D topological insulator possessing TRS and chiral symmetry as
\begin{small}
\be
H_{{\rm DIII}}=\sum_p\Psi^\dagger
\bma
\epsilon_p & 0 & t p_+ & -t p_z \\
0 & \epsilon_p & -t p_z & -t p_- \\
t p_- & -t p_z & -\epsilon_p & 0 \\
-t p_z & -t p_+ & 0 & -\epsilon_p \\
\ema
\Psi,
\ee
\end{small}
where $\Psi(p)=(a_{\uparrow  p},a_{\downarrow p},b_{\uparrow p}, b_{\downarrow p})$, $p_\pm=p_x\pm i p_y$ and $a_{\sigma p}$ and $b_{\sigma p}$ operators indicate the two different sublattices instead of particle-hole degree freedom in $^3$He-B phase. Let $t=2$ and in the lattice model the Hamiltonian can be written as
\be
\mH_{{\rm DIII}}=\epsilon_p\tau_z+\sin p_x \tau_x \siz+\sin p_y \tau_y+\sin p_z \tau_x\six, \label{DIIIR}
\ee
where $\epsilon_p=m-\cos p_x-\cos p_y-\cos p_z$ and $\tau_i/\sigma_j$ describes the sublattice/spin degrees freedom. The expressions of the TRS operator and chiral symmetry operator are $T=\siy\Theta$ and $S=\tau_x\siy\Theta$ so the \emph{pseudo}-PHS operator is $C=\tau_x\Theta$.
Since $T^2=-1$ and $C^2=1$, such a system belongs to class DIII.
Due to chiral symmetry preserving, by using the similar method in sec.\ \ref{windingcompute} the winding number $N_{\bZ}$ can be computed:
\be
N_{\bZ}=
\begin{cases}
-2 &{\rm as}\ |m|<1 \\
1 & {\rm as}\ 1<|m|<3 \\
0 & {\rm elsewhere}
\end{cases}
\ee
In the following, we put our focus on $m=2$. In the original classification, this winding number corresponds to the number of the gapless surface modes and the entanglement mid-gap states, which are intact against any symmetry preserving disorder.

Now we introduce reflection symmetry, which changes $x\rightarrow -x$. There are two possible expressions of the reflection symmetry operators $R$. First, $a_{\sigma p}$ and $b_{\sigma p}$ exchange under reflection so $R=\tau_x\six$ (we require $R$ to be hermitian). This reflection symmetry is broken when the Hamiltonian is in the expression of \Cref{DIIIR}. Secondly, the sublattice is invariant under reflection so $R=\six$. The reflection symmetry is preserved in our system. Therefore, $R$ anticommutes with $T$ and commutes with $C$. (For $^{3}$He-B, such relations change because $R=\tau_z\six$.) Since $R$ commutes with $C$, the Hamiltonian at the reflection symmetry planes $k_x=0,\ \pi$ corresponds to class D in 2 dimensions which has a $\bZ$ topological invariant. Therefore, the mirror Chern numbers can be computed in those symmetry planes. We perform a unitary transformation on the symmetry operators and the Hamiltonian
\be
\six\rightarrow \siz,\ \siz\rightarrow -\six, \Theta\rightarrow \Theta
\ee
so that $R=\siz$ and the Hamiltonian is transformed to
\be
\mH'_{{\rm DIII}}=
\bma
\mH_+ & -\sin p_x \tau_x \\
-\sin p_x & \mH_- \\
\ema,
\ee
where $\mH_\pm=\epsilon_p\tau_z+\sin p_y \tau_y\pm\sin p_z \tau_x$ in the eigenspace of $R=\pm 1$. As $m=2$, by \Cref{evenZ} the Chern numbers for $\mH_+$ at $k_x=0,\ \pi$ are $\nu_0=1, \nu_\pi=0$ respectively. Therefore, the mirror Chern number defined by \Cref{numSZ} is
\be
N_{M\bZ}=1
\ee
The topological phase of the system protected by the $\bZ$ and $M\bZ$ topological invariants is described by the $\bZ^1$ number in \Cref{NbZ1}. In this case, $N_{\bZ^1}=1$ corresponding to the number of the gapless surface states and the entanglement mid-gap.

	Now consider the two copies of the topological insulator Hamiltonian in a system but with a sign changing. The enlarged BdG Hamiltonian can be written as
\be
\mH_{{\rm DIII}}^2=
\bma
\mH_{{\rm DIII}} & 0 \\
0 & \mH_{{\rm DIII}}'  \\
\ema,
\ee
where we choose $\mH_{{\rm DIII}}' =\mH_{{\rm DIII}}(p_x\rightarrow -p_x)$. The two topological numbers for $\mH'_{{\rm DIII}} $ are $N_{M\bZ}=1$ and $N'_{\bZ}=-1$ and then for the whole system are $N_{M\bZ}=2$ and $N_{\bZ}=0$ so $N_{\bZ^1}=2$. Although the $\bZ$ topological number vanishes, the topological phase is protected by $N_{M\bZ}$.

x	Let $\mH_{{\rm DIII}}'$ in $\mH_{{\rm DIII}}^2$ become $\mH_{{\rm DIII}}$ with $p_x\rightarrow -p_x,p_y\rightarrow -p_y$. Although $N_{M\bZ}=0$ for the entire system, $N_{\bZ}=2$ so $N_{\bZ^1}=2$. That is the reason that there are two robust gapless surface modes.

	Also we consider $\mH_{{\rm DIII}}'=\mH_{{\rm DIII}}$ with only $p_y\rightarrow -p_y$. Both of the two topological numbers for $\mH_{{\rm DIII}}'$ switch the signs. Therefore, the total numbers vanish: $N_{M\bZ}$ and $N_{\bZ^1}$. This phase is trivial because we can find a SPEMT
\be
\bma
0 & \tau_y \\
\tau_y & 0 \\
\ema,
\ee
which anticommutes with $\mH_{{\rm DIII}}^2$.

\section{Conclusion}
Combining \Cref{commutationtable,anticommutationtable,gamma1S,gammaleft}, we write down the complete classification results (27 symmetry classes $+R$) in \Cref{table:mirror class}. The classification of reflection symmetric topological insulators and superconductors still has the same spatial dimensional periodicities with the original Altland-Zirnbauer classification in \Cref{originaltable}.  The complex and real symmetry classes with reflection symmetry have the periodsof two and eight respectively.  Although the reflection classification tables seem to be complicated, the two ingredients can slightly simplify the classifications. To define these ingredients, we consider a $d$-dimensional system in an AZ symmetry class with reflection symmetry. We denote $N^d$ as a topological invariant of the original strong index without reflection symmetry as shown in \Cref{originaltable}. Furthermore, we define $N^{d-1}$ as the $(d-1)$-dimensional topological invariant 
in the corresponding mirror symmetry class (appendix \ref{mirror class}). 
By observing the reflection tables, some of the topological invariants of the strong index are determined by $N^d$ and $N^{d-1}$ as shown in \Cref{summarytable}. 
In other cases, there is an ambiguity to determine the strong topological invariants of reflection symmetric systems. Therefore, 
the approach of using minimal Dirac model provides a systematic way to find topological invariants.  

\begin{table}[t]
\begin{center}
\begin{tabular}{|c|c|c|c|}
\hline
\backslashbox{$N^{d-1}$}{$N^d$} & $0$ & $\bZ_2$ & $\bZ$ \\
\hline
0 & 0 & 0, $\bZ_2$ & $0,\ \bZ$ \\
\hline
$\bZ_2$ & $0$ & $\bZ_2$ & $\bZ_2,\bZ$ \\
\hline
$\bZ$ & $0,M\bZ$ & $M\bZ$ & $\bZ^1$ \\
\hline
\end{tabular}
\caption{
The topological invariants of reflection symmetric topological insulators
 and superconductors,
 and
 the strong topological index for $d$-dimensional topological states
 in the original periodic table I ($N_d$),
 and the $d-1$-dimensional mirror topological invariant ($N^{d-1}$).
 In several cases, $N^{d}$ and $N^{d-1}$ still cannot completely determine the reflection topological invariant. We have to go back to the minimal Dirac Hamiltonian method to determine the topological characters.}
\end{center}
\label{summarytable}
\end{table}

While topological insulators and superconductors protected by a set of spatial discrete symmetries 
are more fragile in general than those 
protected by non-spatial symmetries, they are still fairly relevant to realistic systems. We list several realistic reflection symmetric systems in \Cref{table:mirror class}.

For example, we, again, note that
for the $\mathbb{Z}_2$ TRS topological insulator in class AII
(without reflection),
its Dirac representative has reflection symmetry $R_{--}$ in three spatial dimensions.
This is, from the point of view of TRS topological insulators,
somewhat accidental.
However, as we discussed, this is related to the fact that
there is a topological
distinction of ground states even without time-reversal
when reflection symmetry is a good symmetry.
The associated topological invariant is
integer-valued($M\bZ$), as opposed to $\mathbb{Z}_2$.

In fact,
many experimentally realized topological insulators such as Bi$_2$Se$_3$
``accidentally'' have reflection symmetry.
From the discussion above,
even breaking TRS, a surface Dirac cone will not go away
if both the time-reversal symmetry breaking perturbation and the surface geometry respect
reflection symmetry. 
For example, we consider a surface which respects the reflection symmetry. When 
an in-plane magnetic field, namely parallel to the surface, is applied to the system, the time reversal symmetry is broken. However, the reflection symmetry is still preserved when the direction of the in-plane magnetic field is tuned to coincide with the reflected direction. If this occurs, the surface Dirac cone is stable since it is protected by the preserved reflection symmetry even though the time reversal symmetry has been broken. 
These stable surface Dirac cone may be
detected by STM, say.
(ARPES may not be ideal if we use a magnetic field to break TRS). When the direction of applied magnetic field is away from the reflection direction, the system respects neither reflection nor time reversal symmetry. As a consequence, a gap should open in the surface states. We predict that the surface magneto-resistance with an in-plane magnetic field changes significantly, when the field direction is rotated and when the chemical potential is close to the Dirac point. This is because the surface gap varies with the rotation of the field direction. 


We close with discussion on the effects of disorder on 
topological insulators and superconductors protected by
reflection symmetry.  
In the ten-fold classification of topological insulators and superconductors, 
it has been 
proved useful to consider the boundary (edge, surface, etc.) 
Anderson localization problem:
For a topological bulk, one should find 
a boundary mode which is completely immune to disorder.
In turn, once one finds such
``Anderson delocalization'' at the boundary, 
it means there is a topologically non-trivial bulk. 
Not only this bulk-boundary correspondence can be used to 
find and classify bulk topological phases 
{\it in the absence of disorder}, 
it immediately tells us such topological phases are stable
against disorder;
two phases which are topologically distinct cannot be adiabatically
connected by either spatially homogeneous or inhomogeneous deformations.
For topological phases protected by a set of spatial symmetries, 
stability against disorder is, in general, not trivial, since
spatial inhomogeneity does not respect the spatial symmetries. 
One can still consider, however, situations where 
the spatial symmetries are preserved {\it on average}. 
\cite{Yao:2012,Fulga:statTI,Fu:average_symmetry}
Below, we will consider the stability of 
reflection-protected topological phases we identified earlier
against disorder which is reflection symmetric on average. 

Let us consider as an example the reflection symmetric topological insulator
in symmetry class A in three dimensions (class A + $R$).
(For other examples in two dimensions, see Ref.\ \cite{Yao:2012}.)
For symmetry class A in three dimensions,
we have a topological insulator protected by 
reflection symmetry, which is characterized by
an integer topological invariant.
Let us consider the surface Hamiltonian: 
$
\mathcal{H}(r)= \mathcal{H}_0 (r) + \mathcal{V}(r)
$
where $r$ denotes the two-dimensional coordinates
on the surface, $\mathcal{H}_0(r)$ is a kinetic term 
(the surface Dirac kinetic term).
We have added a random perturbation $\mathcal{V}(r)$. 
The disorder-free part is reflection symmetric
under $R$, which is reflection inherited from the bulk,
$
R^{-1} \mathcal{H}_0(\tilde{r}) R
=
\mathcal{H}_0(r),
$
while disorder $\mathcal{V}(r)$ is not so.
The reflection symmetry can be, however, imposed on average:
$
R^{-1}  \overline{\mathcal{V}(\tilde{r})} R
=
\overline{\mathcal{V}(r)},
$
where $\overline{\cdots}$
represents the quenched disorder averaging. 
We could approach this problem by means of effective 
field theories of Anderson localization, the 
non-linear sigma models (NL$\sigma$Ms);
they describe slowly varying degrees of freedom 
in a disordered metal, which are related to 
a diffusion motion of electrons
(called ``diffusions'' and ``Cooperons'').
When derived for the
disordered surface problem,
the action of the NL$\sigma$M is given by 
\begin{align}
S_{\mathrm{NL}\sigma\mathrm{M}}
&=
\frac{1}{\lambda}
\int d^2 r\,
\mathrm{tr}\,  
\left[ \partial_{\mu} Q  \partial_{\mu}Q\right]
\nonumber \\
&\quad 
+
\frac{\Theta}{16\pi {i}}
\int d^2 r\, 
\epsilon^{\mu\nu}
 \mathrm{tr}\,  
\left[ Q \partial_{\mu} Q  \partial_{\nu}Q\right], 
\label{nlsm action}
\end{align}
where
a matrix field $Q(r)$ is 
the NL$\sigma$M field 
$Q\in \mathrm{U}(2N_{{r}})/
\mathrm{U}(N_{{r}})\times \mathrm{U}(N_{{r}})
$
and $N_r$ is the number of replicas;
$\lambda$ the coupling constant of the NL$\sigma$M, 
which is the strength of interactions among diffusons and Cooperons,
and is inversely proportional the conductivity. 
The last term the action is the topological term (Pruisken term)
,which counts the non-trivial winding 
associated to 
$\pi_2 [\mathrm{U}(2N_{{r}})/\mathrm{U}(N_{{r}})
\times \mathrm{U}(N_{{r}})]=\mathbb{Z}$. 
(Here we are considering the real space which is topologically equivalent
to a sphere.) 
In the absence of any discrete symmetry,
$\Theta$ can take, in principle, any value, 
in which case,
one can make electrons to be Anderson localized. 
While the action for generic value of $\Theta$ breaks reflection symmetry
(on the surface),
$\Theta = (\mbox{integer}) \times \pi$ turns out to be consistent
with reflection symmetry.
Moreover, when $\Theta = (\mbox{odd integer}) \times \pi$ 
there is no Anderson localization. 
The NL$\sigma$M (\ref{nlsm action})
can be derived for 
the Dirac representative of the surface mode,
which consists of $N$ flavors of two-component Dirac fermions
when the bulk topological invariant is $N\in \mathbb{Z}$. 
The theta angle is given by $\Theta = N \pi$. 
Therefore there is an even odd effect;
when the bulk topological invariant is odd (even), 
the surface mode is stable (unstable) against disorder.
This would mean that, in the presence of spatially inhomogeneity 
which nevertheless preserves reflection symmetry on average, 
the topological distinction is not $\mathbb{Z}$, but $\mathbb{Z}_2$.

\acknowledgments

We would like to thank
Jens Bardarson,
Po-Yao Chang, 
Chen Fang,
Akira Furusaki,
Taylor Hughes,
Steve Kivelson,
Dung-Hai Lee,
Christopher Mudry,
Xiao-Liang Qi,
Michael Stone,
Jeffrey Teo,
and
Shou-Cheng Zhang,
for helpful discussions.
CKC is supported by the NSF under grant DMR 09-03291. HY is partly supported by Tsinghua Startup Funds and the National Thousand Young Talents Program. 
\\

While we were finalizing the paper, a preprint\cite{Sato:classD}
appeared in which symmetry class D with reflection symmetry
is discussed. 
One of their considerations corresponds to class D + $R_{++}$ in our general
classification scheme.

\appendix

\section{The calculation of $\bZ$ number}	\label{Zcal}
	We review the calculation of the topological number for the symmetry classes possessing a $\bZ$ or $2\bZ$ topological invariant. For the classification without reflection symmetry, in odd ($d=2n+1$) spatial dimensions such symmetry classes always have chiral symmetry, which is a key point to define the topological number. In even ($2n+2$) spatial dimensions, the topological number is exactly the same with the $n+1$-th Chern number.
		
	To seek the expression of the topological number in $2n+1$ spatial dimensions, first we introduce the spectral projector onto the filled Bloch states and the ``Q-matrix'' (flat band) \cite{SRFLnewJphys} by
\be
P(k)=\sum_{\hat{a}} \ket{u^{-}_{\hat{a}}(k)} \bra{u^-_{\hat{a}}(k)},\ Q(k)=\bI-2P(k).
\ee
We note that $P(k)^2=P(k)$ so $Q(k)^2=\bI$. If the system possesses the chiral symmetry, which is described by \cref{chiral}, with negative energy filled $N_+=N_-$. The Q-matrix can be brought into block off-diagonal form,
\be
Q(k)=
\begin{pmatrix}
0 & q(k) \\
q^\dagger (k) & 0
\end{pmatrix},\ q(k)\in U(N_-). \label{Qmatrix}
\ee
in some basis. The topological number in $2n+1$ spatial dimensions is characterized by the winding number
\be
\nu_{2n+1}[q]=\frac{(-1)^nn!}{(2n+1)!}(\frac{i}{2\pi})^{n+1}\int_{\mathrm{BZ}^{d=2n+1}}\mathrm{tr}[(q^{-1}dq)^{2n+1}] \label{oddZ}
\ee
Here, BZ$^d$ means the integration over $d$-dimensional $k$-space. That is, the region is the first Brillouin zone in the lattice model.

For the topological number in $2n+2$ spatial dimensions, first we define the non-Abelian Berry connection of the occupied bands
\be
\mathcal{A}^{\hat{a}\hat{b}}(k)=A^{\hat{a}\hat{b}}_\mu(k)dk_\mu=\braket{u_{\hat{a}}^-(k)}{du^-_{\hat{b}}(k)},
\ee
where $\mu=1,\cdots,d,\ \hat{a},\ \hat{b}=1,\cdots,N_-$. And the Berry curvature is defined by
\be
\mathcal{F}^{\hat{a}\hat{b}}(k)=d\mathcal{A}^{\hat{a}\hat{b}}+(\mathcal{A}^2)^{\hat{a}\hat{b}}
\ee
The topological number is captured by the $n+1$-th Chern number
\be
\nu_{2n+2}=\mathrm{Ch}_{n+1}[\mathcal{F}]=\frac{1}{(n+1)!}\int_{\mathrm{BZ}^{d=2n+2}}\mathrm{tr}(\frac{i\mathcal{F}}{2\pi})^{n+1} \label{evenZ}
\ee


\section{Symmetry classes in mirror planes}\label{mirror class}
In the mirror planes $k_1=0,\ \pi$, the Hamiltonian commutes with the reflection operator $R$
\be
[\mH_{k_1=0,\pi},R]=0.
\ee
Furthermore, because $R$ is hermitian and $R^2=1$, the Hamiltonian can be decomposed to two diagonal blocks $\mH_{\pm}$ corresponding to the eigenspaces $R=\pm 1$ respectively. Now consider \emph{non-spatial} symmetries, which determine the symmetry class of the system. If all of the \emph{non-spatial} symmetry operators commute with $R$, the two block Hamiltonians $\mH_{\pm}$ belong to the same symmetry class. Furthermore, we name this symmetry class for $\mH_{\pm}$ in the mirror planes as \emph{mirror symmetry class}. However, the mirror symmetry class changes when at least one of the \emph{non-spatial} symmetry operators anticommutes with $R$. The Hamiltonians $\mH_{\pm}$ are invariant under the $R$-commuting symmetries but not invariant under the $R$-anticommuting symmetry. Therefore, the mirror symmetry class is determined by the $R$-commuting symmetries. \Cref{table:mirror class} shows the mirror symmetry classes for each possible algebraic relation between \emph{non-spatial} symmetry operators and $R$.

\section{The proof of the $\bZ^1$ number definition }\label{bZ1proof}
We will prove that the bulk topology of the $\bZ^1$ system is determined by the maximum value of $|N_{\bZ}|$ and $|N_{M\bZ}|$ in general cases. We simplify the problem by considering the Dirac Hamiltonian with the coefficient of the mass term:
\be
m=M-(k_x\pm \delta_i)^2-\tilde{k}^2.
\ee
In the proof of sec.\ \ref{bZ1} we only consider $\delta_i=0$. In general, $\delta_i$ can be any number in BZ. As $\delta_i=0, \pi$, $N_{\bZ}$ and $N_{M\bZ}$ can be computed for such Dirac Hamiltonians. As $\delta_i$ is not at the mirror symmetry points, the Dirac Hamiltonians only make a contribution to $N_{\bZ}$. To achieve the proof of  the $\bZ^1$ number definition, we consider the Dirac Hamiltonians in any possible distribution and determine the interplay between the phase of the non-trivial Dirac Hamiltonians, $N_{\bZ}$, and $N_{M\bZ}.$

In general, the $M\bZ$ topological number is computed in the two symmetry planes $k_x=0, \pi$.
To simplify the problem, we consider the weak mirror index in \Cref{mirror weak} vanishes so $\nu_0\nu_\pi\leq 0$; hence, the $M\bZ$ number is given by $N_{M\bZ}=\nu_0+\nu_\pi$.
However, with translational symmetry breaking by folding the BZ, the two symmetry planes collapse to the one symmetry plane $k=0$.
The $M\bZ$ number of this $d-1$-dimensional plane is $\nu_0+\nu_\pi$.
Since $N_{M\bZ}$ is invariant under density waves that connect these two points, the $d-1$-dimensional $\bZ$ number in the new $k_x=\pi$ plane vanishes. Thus, we still can discuss any $M\bZ$ number only in the $k_x=0$ plane without loss of generality.

	 The $\bZ$ topological number ($N_{\bZ}$) for the entire system, in general, cannot be defined in the mirror symmetry planes. The discussion of the $\bZ$ number can be separated to two parts: $\delta_i=0, \pi$ and $\delta_i\neq 0$ or $\pi$. First, the part $N_{\bZ}^0$ of the $\bZ$ number is contributed from the Dirac Hamiltonians with $\delta_i=0,\ \pi$. The problem can be simplified by considering only $\delta_i=0$ case. The reason is that the contribution of the $\delta_i=\pi$ can be moved to $\delta_i=0$ by folding the BZ. Secondly, the other part $N_{\bZ}'$ of the $\bZ$ number are contributed from $\delta_i\neq 0$ or $\pi$. Due to the reflection symmetry, $N_{\bZ}'$ must be even. Out of the mirror symmetry planes, the bulk topology is losing the reflection symmetry protection. Therefore, if the Dirac Hamiltonians ($\pm\delta_i$) possess $\bZ$ numbers that differ by signs, the system combined by these two Hamiltonian is in the trivial phase. Hence, we put our focus on these Dirac Hamiltonians having the same sign of the $\bZ$ numbers.


	 We define $(+)$ as a pair of the Dirac Hamiltonian ($\pm\delta_i$) with an even number of the gamma matrices having the minus sign. In other words, the $\bZ$ number in this case is $2$. Similarly, $(-)$ indicates an odd number of the minus-sign gamma matrices and $N_{\bZ}'=-2$. Consider a system possessing the Dirac Hamiltonians with $(+,+)$ (see sec.\ \ref{bZ1}) and $(-)$. Hence, $N_{\bZ}=-1$ and $N_{M\bZ}=1$. The entire Hamiltonian for the system is written as
\begin{align}
H=&\sum_{k_1}[\mH_{++}b^\dagger_{k_1}b_{k_1}
+\mH_{-}(c^\dagger_{k_1+\delta}c_{k_1+\delta}+c^\dagger_{k_1-\delta}c_{k_1-\delta})],
\end{align}	
\noindent where $\mH_{++}=m\gamma_0+\sum_{i=1}k_i\gamma_1$ and $\mH_{-}=\sum_{n_i}k_{n_i}\gamma_{n_i}-\sum_{n_j}^{{\rm odd}}k_{n_j}\gamma_{n_j}$. Due to \Cref{Rgamma} the Hamiltonian is invariant under the reflection symmetry with the reflection operator in the second quantization
\be
\hat{R}=\sum_{k_1}(b_{k_1}^\dagger Rb_{-k_1}+c_{k_1}^\dagger Rc_{-k_1}).
\ee
The SPEMT is found to prevent two of the three minimal Dirac Hamiltonians passing through quantum phase transition: 	
\be
\hat{\mN}_{\pm}=\frac{\mN}{\sqrt{2}} (b^\dagger_{k_1}c_{k_1+\delta}\pm b^\dagger_{k_1}c_{k_1-\delta})+h.c.,
\ee
where $\mN=(i)\prod_{n_j}^{{\rm odd}}\gamma_{n_j}$ and we choose the presence or absence of ``$i$'' to preserve TRS and PHS. Moreover, in order to preserve the reflection symmetry $+/-$ in $\hat{\mN}$ corresponds to $n_j\neq 1/n_j= 1$ in the product of $\mN$. The entire Hamiltonian with $\hat{\mN}_{\pm}$ and a real coupling coefficient $g$ is given by
\be
H_{\pm}=\sum_{k_1}\Psi^\dagger_{\pm} \bma
\mH_{++} & g\mN & 0 \\
g\mN^\dagger & \mH_{-} & 0 \\
0 &  0 & \mH_{-} \\
\ema\Psi_{\pm}
\ee
in the basis of
\be
\Psi_{\pm} =
\bma
c_{k_1} & \frac{1}{\sqrt{2}}(c_{k_1+\delta}\pm c_{k_1-\delta})  & \frac{1}{\sqrt{2}}(c_{k_1+\delta}\mp c_{k_1-\delta})
\ema ^T.
\ee
Hence, the first two Dirac Hamiltonian become trivial and the last Dirac Hamiltonian survives. The topological number is one, which equals to $N_{\bZ^1}$ in \Cref{NbZ1}.

Now we add an extra $(+,+)$ into the original system. That is, $N_{\bZ}=0$ and $N_{m\bZ}=2$. The entire Hamiltonian with the SPEMTs is written as
\be
H_{\pm}=\sum_{k_1}\Phi^\dagger_{\pm} \bma
\mH_{++} & 0 & g_1\mN & 0 \\
0 & \mH_{++} & g_2\mN & 0 \\
g_1\mN^\dagger & g_2\mN^\dagger & \mH_{-} & 0 \\
0 &0 &  0 & \mH_{-} \\
\ema\Phi_{\pm}
\ee
in the basis of
\be
\Phi_{\pm} =
\bma
a_{k_1} & b_{k_1} & \frac{1}{\sqrt{2}}(c_{k_1+\delta}\pm c_{k_1-\delta})  & \frac{1}{\sqrt{2}}(c_{k_1+\delta}\mp c_{k_1-\delta})
\ema ^T.
\ee
The SPEMT coupling from the last Dirac Hamiltonian $\mH_{{\rm odd}}$ to the others is forbidden by the symmetries. By performing a proper unitary transition, only two Dirac Hamiltonian couple so that these subsystems become trivial. The two other Dirac Hamiltonian keep the original bulk topology structure. Hence, the topological number is two, which is the maximum value of $|N_{\bZ}|$ and $|N_{M\bZ}|$.

Instead of adding an extra ($+,+$), we add ($-,-$)
\be
\mH_{--}=-m\gamma_0-k_1\gamma_1+\sum_{i=2} k_i\gamma_i
\ee
into the original system. In other words, consider the system with $N_{\bZ}=N_{M\bZ}=0$. The entire Hamiltonian with the SPEMT is given by
\be
H_{\pm}=\sum_{k_1}\Phi^\dagger_{\pm} \bma
\mH_{--} & 0 & 0 & g_1\frak{L} \\
0 & \mH_{++} & g_2\mN & 0 \\
0 & g_2\mN^\dagger & \mH_{-} & 0 \\
g_1\frak{L}^\dagger &0 &  0 & \mH_{-} \\
\ema\Phi_{\pm},
\ee
where $\frak{L}=i\mN \gamma_0 \gamma_1$ preserves all of the symmetries. All of the Dirac Hamiltonian couple so the entire system is in the trivial phase. As expected, the topological number vanishes. Thus, only one ($+,+$) can couple with a pair Dirac Hamiltonian ($-$) and then one Dirac Hamiltonian survives and provides the non-trivial bulk topology. In general, consider $N_{+,+}$ ($+,+$) Dirac Hamiltonians and $N_{-}/2$ pairs of $(-)$ Dirac Hamiltonians in a system. The number of the nontrivial Dirac Hamiltonians is $|N_{+,+}-N_{-}/2|$. Similarly, the number of the protected modes for the other cases separately are $|N_{+,-}-N_{+}/2|$, $|N_{-,-}-N_{-}/2|$, and $|N_{-,+}-N_{+}/2|$, where $N_{+}/2$ is the number of the pairs ($+$) Dirac Hamiltonians. However, the discussion in sec.\ \ref{bZ1} $(+,\pm)$ the number of the protected modes for the entire system is $|N_{+,+}-N_{+,-}|+|N_{-,+}-N_{-,-}|$. In this case, we replace
\begin{align}
N_{\pm,\mp}\rightarrow N_{\pm,\mp}-N_{+}/2 \nonumber \\
N_{\pm,\pm}\rightarrow N_{\pm,\mp}-N_{-}/2
\end{align}
Moreover, a system with the same number of ($-$) and $(+)$ is trivial. Thus, the total number of the protected modes is given by
\be
|N_{+,+}-N_{+,-}+\frac{N_+-N_-}{2}|+|N_{-,+}-N_{-,-}+\frac{N_+-N_-}{2}|,
\ee
which is the topological number for the $\bZ^1$ system. By applying those identities
\begin{align}
N_{\bZ}^0\pm N_{M\bZ}=N_{\pm,+}-N_{\pm,-}, \\
N_{\bZ}'=N_+-N_-,\ N_{\bZ}=N_{\bZ}^0+N_{\bZ}',
\end{align}
we can simplify the expression of the $\bZ^1$ number:
\begin{align}
N_{\bZ^1}=&|\frac{N_{\bZ}^0+N_{M\bZ}+N_{\bZ}'}{2}|+|\frac{N_{\bZ}^0-N_{M\bZ}+N_{\bZ}'}{2}| \nonumber \\
=&{{\rm Max}}(|\nu_d|,|N_{M\bZ}|).
\end{align}

	

	


\bibliographystyle{apsrev4-1}
\bibliography{TOPO3}

\begin{thebibliography}{53}
\bibitem [{gam()}]{gamma_dplus3}%
  \BibitemOpen
  \href@noop {} {}\bibinfo {howpublished} {We give two examples: for class CII
  in $d=2$, the symmetry operators are $C=\tau_x\mu_y$ and $T=\sigma_x\Theta$.
  The gamma matrices are $\gamma_0=\tau_z,\ \gamma_1=\tau_x\sigma_x,\
  \gamma_2=\tau_x\sigma_y\mu_x,\ \gamma_3=\tau_x\sigma_z,\ \gamma_4=\tau_y,$
  and $\gamma_5=\tau_x\sigma_y\mu_z$. For class C in $d=3$, only PHS is kept
  and the gamma matrices are the same with class CII, except for the extra
  kinetic term $\gamma_6=\tau_x\sigma_y\mu_y$.}\BibitemShut {Stop}%
\bibitem [{oth()}]{other:reflection}%
  \BibitemOpen
  \href@noop {} {}\bibinfo {howpublished} {The other definition of the
  reflection symmetry is
  $\hat{R}=\sum_{k_1}c^\dagger_{k_1}(i\gamma_0\gamma_{d+1})c_{-k_1}e^{-ik_1}$.
  We redefine $\mN=e^{-i\eta}i\gamma_{d+1}\gamma_{d+2}\gamma_{d+3}$. The
  topological phase is still trivial.}\BibitemShut {Stop}%
\end{thebibliography}

\end{document}